\documentclass[%
 reprint,
superscriptaddress,
 amsmath,amssymb,
 aps,
pra,
floatfix,
]{revtex4-2}
\usepackage{graphicx}% Include figure files
\usepackage{dcolumn}% Align table columns on decimal point
\usepackage{bm}% bold math
\usepackage{mathtools}
\usepackage{color}
\usepackage{physics}
\usepackage{comment}
\usepackage{hyperref}% add hypertext capabilities
\usepackage{booktabs}

\providecommand{\ket}[1]{|#1\rangle}

\begin{document}

\title{Gradient Ascent Pulse Engineering for Multi-Time Correlations}% Force line breaks with \\

\author{Issei Matsuzoe}
\thanks{These authors contributed equally to this work.}
\affiliation{%
Department of Electronics and Electrical Engineering, Faculty of Science and Technology, Keio University, Yokohama, Kanagawa 223-8522, Japan
}%

\author{Kazufumi Tanji}
\thanks{These authors contributed equally to this work.}
\affiliation{%
Department of Electronics and Electrical Engineering, Faculty of Science and Technology, Keio University, Yokohama, Kanagawa 223-8522, Japan
}%

\author{Wojciech Roga}
\affiliation{%
 Department of Electronics and Electrical Engineering, Faculty of Science and Technology, Keio University, Yokohama, Kanagawa 223-8522, Japan
}%

\author{Masahiro Takeoka}
\affiliation{%
 Department of Electronics and Electrical Engineering, Faculty of Science and Technology, Keio University, Yokohama, Kanagawa 223-8522, Japan
}%
\affiliation{Advanced ICT Research Institute, National Institute of Information and
Communications Technology (NICT), Koganei, Tokyo 184-8795, Japan}
\date{November 16, 2026}

\begin{abstract}
  In quantum optimal control, gradient methods are often used to optimize functions of a state or propagator at a final time, while experimentally addressable observables and performance metrics may depend on multi-time correlations. We develop QRT-GRAPE, which combines the quantum regression theorem (QRT) with Gradient Ascent Pulse Engineering (GRAPE) to obtain analytic gradients for objective functions defined by multi-time correlations. The method incorporates the time evolution of an auxiliary operator in QRT into the forward propagation and backward adjoint propagation used in GRAPE. As an application, we consider the trade-off between the photon-emission probability and the conditional Bell-state fidelity in heralded remote entanglement generation among two nonidentical atom--cavity systems. Starting from a flat-top Rabi pulse pair selected by a search over pulse amplitudes and durations, QRT-GRAPE increases the photon-emission probability product from $0.2802$ to $0.3216$ and the conditional Bell-state fidelity from $0.9356$ to $0.9441$.
\end{abstract}

\maketitle

\section{\label{sec:1} Introduction}

The correlation functions allow us to analyze statistical relations between quantities and events that do not necessarily occur in the same times. They are widely used in different disciplines, especially in which the relations and dynamics are determined by statistical laws. In quantum physics, they become fundamental to connect quantum dynamics and experimentally accessible observables in different times and to calculate operationally important functions such as fidelity between quantum states or quantum entanglement.  
%Correlation functions connect quantum dynamics to experimentally accessible observables.
They define spectroscopic and linear-response signals~\cite{Dorfman2016Nonlinear,Kubo1957StatisticalMechanical}, determine dynamical structure factors used to characterize quantum many-body systems~\cite{Baez2020Dynamical,Boothroyd2020Principles}, %Beyond characterization of single photons, correlation functions are also used 
and play a role in principal component analysis of continuous variable quantum states~\cite{Takase2019Complete,Kiilerich2019Input}.

In quantum optics, the correlation functions characterize single-photon sources. The second-order coherence function, $g^{(2)}(\tau)$, measures photon statistics and antibunching~\cite{Glauber1963Quantum,Gardiner2004Quantum,Walls2008Quantum,Kimble1977Photon,Fischer2016Dynamical}, whereas the unnormalized first-order coherence function, $G^{(1)}(t,t')$, captures temporal coherence and  temporal mode structure.
Temporal mode overlap is central to two-photon interference and photon indistinguishability~\cite{Fischer2016Dynamical,Walker2020Improving,Kikura2025Highpurity,Tanji2024Ratefidelity}, which directly characterize fidelities of atom--photon entanglements~\cite{Appel2022Entangling,Bracht2024Theory} and remote entangled states generated by interferometric measurements~\cite{Duan2001Longdistance,Sangouard2011Quantum,Hofmann2012Heralded,Bernien2013Heralded,Krutyanskiy2023Entanglement,vanLeent2022Entangling,Tanji2024Ratefidelity,Tanji2026High,aoki2026remote}.
%Rather than serving purely as diagnostic tools,  these correlation functions act as direct performance metrics linking local photon sources to larger photonic protocols, since a remote entangled state, a fundamental building block of the protocols, is evaluated by the functions.

Since the states evaluated by correlation functions form fundamental building blocks in quantum information processing, the correlation functions are directly related to the performance of protocols of quantum technology.
For example, photonic links in quantum networks and distributed quantum architectures rely on the correlation properties of propagating photons~\cite{Nickerson2013Topological,Monroe2014Largescale,Nickerson2014Freely,Li2024HighRate,Sunami2025Scalable,Wehner2018Quantum,Azuma2023Quantum,Sinclair2025Faulttolerant,Singh2025Modular,Singh2026Faulttolerant,Kikura2026Passive}.
Moreover, correlation-based performance quantities also arise in distributed quantum sensing based on entanglement~\cite{Liu2021Distributed,Ueda2025Quantum}.

Specific values of correlation functions may be desired between selected observables in particular times. As such, a correlation function becomes the figure of merit in designing controlled dynamics leading the system from the initial state to the target configuration. This is the subject of optimal control theory. In this case, its goal is to  
%This motivates 
optimize control pulses using objective functions defined by correlation functions.

\begin{figure*}
  \centering
  \includegraphics[width=\textwidth]{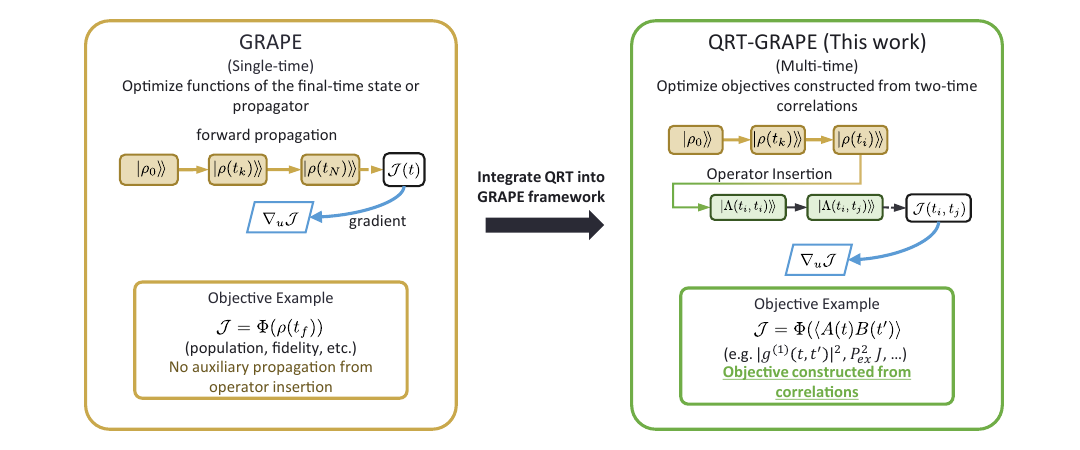}
  \caption{Conceptual comparison between GRAPE for objective functions depending on the state or propagator at the final time and QRT-GRAPE for objective functions constructed from correlation functions.
  In GRAPE, the state is propagated forward in time and the adjoint variable backward in time, without auxiliary propagation initiated by an operator insertion.
  In QRT-GRAPE, the operator insertion required by the quantum regression theorem starts auxiliary propagation for pairs of times $t_i\leq t_j$.
  This construction accommodates objective functions constructed from correlation functions, including measures of photon indistinguishability and temporal mode matching in open quantum systems.}
  \label{fig:qrt-grape}
\end{figure*}

Quantum optimal control (QOC) provides multiple methods for designing time-dependent control pulses to achieve desired dynamical objectives~\cite{Glaser2015Training,Koch2022Quantum}.
Among them, Gradient Ascent Pulse Engineering (GRAPE)~\cite{Khaneja2005Optimal} is a widely used powerful %gradient-based 
approach.
GRAPE and its variants have been applied to state preparation~\cite{Chen2022Iterative,Lu2024Optimal,Khaleghian2026Development}, gate optimization~\cite{Khaneja2005Optimal,Lewis2025Quantum,Malarchick2026When}, quantum sensing~\cite{Li2017Hybrid,Dong2020Learning}, and open-system control~\cite{Petruhanov2023Incoherent,Petruhanov2023GRAPE,Chen2025Robust}.
%Gradient evaluation reuses the states or propagators computed forward in time and the adjoint variables computed backward in time.
However, as shown in Fig.~\ref{fig:qrt-grape}, standard GRAPE is primarily formulated for objective functions evaluated at a single final time, such as state or gate fidelity.
In contrast, correlation functions depend on multi-time observables, to which standard GRAPE methods are not applicable. The goal of this work is to adapt the GRAPE method to correlation function targets.

%Our approach rely on the quantum regression theorem (QRT) which provides a route 

Multi-time correlations such as a first-order coherence function $\langle\hat a^\dagger(t)\hat a(t')\rangle$~\cite{Lax1963Formal,Guarnieri2014Quantum,Lonigro2022Quantum} are the subject of quantum regression theorem (QRT).
For objective functions constructed from multi-time correlations, the strategy shown in Fig.~\ref{fig:qrt-grape} includes an additional step beyond the standard time evolution: an operator corresponding to the correlator is inserted at the earlier time, and the resulting auxiliary operator is then propagated to evaluate the correlation function. Therefore, the standard GRAPE is not directly applicable to these correlation functions.

In this paper, we develop \emph{QRT-GRAPE}, which combines QRT with GRAPE to optimize multi-time correlation functions.
When the requirements of the QRT are satisfied, the auxiliary operator evolves with the same time-dependent generator as the density operator.
We use the common time-dependent generator governing the density operator and the QRT auxiliary operator to incorporate the operator insertion and auxiliary propagation into GRAPE.
This yields analytic gradients for objective functions constructed from multi-time correlations and opens the GRAPE-type methods to the new class of objective functions.

As a proof-of-concept application, we optimize two atom--cavity single-photon sources for heralded remote entanglement generation, which has been demonstrated experimentally~\cite{Krutyanskiy2023Entanglement}. For distributed quantum information processing, both the entanglement generation probability and fidelity are crucial metrics, but they are subject to a trade-off~\cite{Tanji2024Ratefidelity}. Starting from Rabi pulse pairs selected by parameter sweeps, QRT-GRAPE optimizes the Rabi frequencies and increases both the photon-emission probability product and the conditional Bell-state fidelity.
This result indicates that our method can overcome the fundamental bottleneck in quantum networks, namely the trade-off between the photon-emission probability and the conditional Bell-state fidelity. In practical networks, the parameters of the two atom--cavity systems can differ~\cite{Kikura2026Passive,Nolleke2013Efficient}. We therefore investigate the case of connecting two nonidentical cavities, in which we have swept the atom--cavity coupling ratio $g_B/g_A$ and the external cavity decay rate ratio $\kappa_{\mathrm{ex},B}/\kappa_{\mathrm{ex},A}^{\mathrm{opt}}$. At all pairs of ratios, QRT-GRAPE increased the objective function value $\mathcal{J}$, the photon-emission probability product $P_{\mathrm{ex},A}P_{\mathrm{ex},B}$, and the Bell-state fidelity $F_{\mathrm{Bell}}$ relative to the values obtained with the Gaussian Rabi pulse pair selected at the same pair of ratios.
In this sweep, the results suggest that weak atom--cavity coupling limits the optimized performance more strongly than the external cavity decay rate mismatch.

This paper is organized as follows.
Section~\ref{sec:2} reviews GRAPE, the Lindblad master equation, and the QRT.
Section~\ref{sec:3} derives the QRT-GRAPE gradient formula.
Section~\ref{sec:4} presents the cavity QED application and numerical results.
Section~\ref{sec:5} concludes with a discussion of broader applicability and future directions.

%==========================section 2==========================
\section{\label{sec:2} Preliminaries}
%==========================sub-section 2.1==========================
\subsection{\label{sec:2.1} GRAPE}
GRAPE~\cite{Khaneja2005Optimal} is a widely used gradient-based quantum-control method. We summarize the forward--backward structure needed for the QRT-GRAPE extension, following Ref.~\cite{Ansel2024Introduction}. For simplicity, we explain the case of pure states, while it can be easily extended for the mixed state with vectorization of matrices, as we describe later.

We consider a quantum system governed by the Hamiltonian
\begin{equation}
  \hat{\mathcal H}(t)
  =
  \hat{H}_0+
  \sum_{m=1}^M u_m(t)\hat{H}_m,
  \qquad
  u_m(t)\in\mathbb{R},
\end{equation}
where $\hat H_0$ is the static Hamiltonian and $\hat H_m$ is the control Hamiltonian associated with the real control function $u_m(t)$. GRAPE partitions the total evolution time $t_f$ into $N$ slices of duration $\Delta t=t_f/N$ and takes
\begin{equation}
  u_m(\tau_k)=u_{m,k},
  \qquad
  \tau_k\in[t_{k-1},t_k),
\end{equation}
where $1\leq k \leq N$.
With $\hat{\mathcal H}_k=\hat H_0+\sum_m u_{m,k}\hat H_m$, the $k$th sliced propagator is $U_k=e^{-i\Delta t\hat{\mathcal H}_k}$, so that the final state is
\begin{equation}\label{eq:grape_prop}
\ket{\psi(t_f)} = \hat U_N \hat U_{N-1} \cdots \hat U_1 \ket{\psi_0}.
\end{equation}

For a real-valued objective function $\mathcal J=G(\ket{\psi_N})$ at the final time, with $\ket{\psi_N}\equiv\ket{\psi(t_f)}$, we define the terminal adjoint state $\bra{\chi_N}$ as 
\begin{equation}
  \bra{\chi_N}
  =
  \frac{\partial G}{\partial \ket{\psi_N}},
\end{equation}
The derivative $\partial G/\partial\ket{\psi_N}$ is taken with $\bra{\psi_N}$ held fixed.
Since $G$ is real, the two conjugate contributions to its first-order variation give
\begin{equation}
\begin{aligned}
\delta G
&=\langle\chi_N|\delta\psi_N\rangle
+\langle\delta\psi_N|\chi_N\rangle\\
&=2\,\mathrm{Re}\,\langle\chi_N|\delta\psi_N\rangle,
\end{aligned}
\end{equation}
where $\ket{\delta\psi_N}$ is the first-order change in the terminal state induced by variations of the control values.
For a change $\delta u_{m,k}$ in a single control value, the first-order change in $U_k$ is $(\partial U_k/\partial u_{m,k})\delta u_{m,k}$.
Since only $U_k$ depends on this control value, Eq.~\eqref{eq:grape_prop} gives
\begin{equation}
\ket{\delta\psi_N}
=U_N\cdots U_{k+1}
\frac{\partial U_k}{\partial u_{m,k}}
\ket{\psi_{k-1}}\,\delta u_{m,k}.
\end{equation}
Substituting this expression into the first-order variation of $G$ and using $\delta G=(\partial G/\partial u_{m,k})\delta u_{m,k}$ gives
\begin{equation}
\frac{\partial G}{\partial u_{m,k}}
=2\,\mathrm{Re}\left[
\bra{\chi_k}
\frac{\partial U_k}{\partial u_{m,k}}
\ket{\psi_{k-1}}
\right],
\end{equation}
where $\ket{\psi_{k-1}}$ and $\bra{\chi_k}$ are given by forward and backward propagations:
\begin{equation}
    \begin{aligned}
        \ket{\psi_{k-1}}=&U_{k-1}\cdots U_1\ket{\psi_0},\\
        \bra{\chi_{k}}=&\bra{\chi_N}U_N\cdots U_{k+1},
    \end{aligned}
\end{equation}
respectively.
The derivative is evaluated by
\begin{equation}\label{eq:slice_prop_derivative}
    \frac{\partial U_k}{\partial u_{m,k}}
  =-i\Delta t\int_0^1 e^{-i(1-s)\Delta t\hat{\mathcal H}_k}\hat H_m e^{-is\Delta t\hat{\mathcal H}_k}\,ds
\end{equation}
For sufficiently small $\Delta t$, the slice derivative reduces to the usual GRAPE approximation,
\begin{equation}
  \frac{\partial U_k}{\partial u_{m,k}}
  \simeq
  -i\Delta t\,\hat H_m U_k.
\end{equation}
Substituting this relation gives the gradient:
\begin{equation}\label{eq:grape_grad}
  \frac{\partial G}{\partial u_{m,k}}
  \simeq
  2\Delta t\,\mathrm{Im}
  \left[
  \bra{\chi_k}
  \hat H_m
  \ket{\psi_k}
  \right].
\end{equation}
The partial derivatives of the objective function for all $(m,k)$ can therefore be evaluated from the states $\ket{\psi_k}$ and adjoint states $\bra{\chi_k}$ obtained in one forward and one backward propagation, respectively.
Conversely, the na\"{i}ve finite-difference method requires to solve the full propagations iteratively for each partial derivative. Thus, the required number of propagations scales as $NM$ for finite differences, whereas GRAPE requires only two: one forward and one backward propagation.

\subsection{\label{sec:2.2} Quantum Regression Theorem}
For a Markovian open quantum system, the dynamics are governed by the Lindblad master equation~\cite{Lindblad1976On,Gorini1976Completely}:
\begin{align}
    \label{eq:lindblad_general}
    \frac{d}{dt}\rho(t) &= \mathcal{L}(t)\,\rho(t) \nonumber \\
    &= -i[\hat{\mathcal H}(t),\,\rho(t)]
    + \sum_j \left(
        \hat{L}_j\,\rho(t)\,\hat{L}_j^\dagger
        - \tfrac{1}{2}\{\hat{L}_j^\dagger \hat{L}_j,\,\rho(t)\}
    \right),
\end{align}
where $\mathcal L(t)$ is the Liouvillian superoperator, $\hat{\mathcal H}(t)$ denotes the system Hamiltonian, and the Lindblad operators $\{\hat L_j\}$ describe dissipation. 

Expectation values are given by $\langle\hat A(t)\rangle=\mathrm{Tr}[\hat A\rho(t)]$, while the density matrix cannot determine multi-time correlations such as the first-order correlation $G^{(1)}(t,t+\tau)=\langle\hat a^\dagger(t)\hat a(t+\tau)\rangle$. Computing such correlations therefore requires an additional relation.

The quantum regression theorem (QRT), introduced by Lax~\cite{Lax1963Formal}, provides a way to obtain multi-time correlations through auxiliary propagation.
For simplicity, we first consider a two-time correlation and later generalize the construction to multi-time correlations. For $\tau\ge0$, we consider the correlator $\langle\hat A(t)\hat B(t+\tau)\rangle$. According to QRT, the correlator is represented as
\begin{equation}\label{eq:two_time_corr_haisenberg}
  \langle \hat{A}(t)\,\hat{B}(t+\tau)\rangle
  = \mathrm{Tr}\!\bigl[\hat{B}\;\hat \Lambda(t,\,t+\tau)\bigr],
\end{equation}
where the auxiliary operator $\hat \Lambda(t,t+\tau)$ evolves from $t$ to $t+\tau$ with the same generator as the Lindblad master equation,
\begin{align}\label{eq:qrt}
  \frac{\partial}{\partial\tau}\hat\Lambda(t,\,t+\tau)
  &= \mathcal{L}(t+\tau)\;\hat\Lambda(t,\,t+\tau),\\
  \hat \Lambda(t,t)&=\rho(t)\hat A.\nonumber
\end{align}
For the oppositely ordered correlator, $\langle\hat A(t+\tau)\hat B(t)\rangle$, the insertion of $\hat B$ gives another auxiliary operator $\hat{{\Lambda}}^\prime(t,t)=\hat B\rho(t)$, and the correlator is evaluated with $\hat A$ at $t+\tau$:
\begin{equation}
  \langle \hat{A}(t+\tau)\hat{B}(t)\rangle
  =
  \mathrm{Tr}\!\left[
    \hat{A}\hat{\Lambda}^\prime(t,t+\tau)
  \right].
  \label{eq:qrt_opposite_order}
\end{equation}
The subsequent auxiliary propagation uses the same time-dependent Liouvillian as in Eq.~\eqref{eq:qrt}.

A time-ordered multi-time correlator,
\begin{equation}
    C(t_1,\ldots,t_N)=\langle \hat A_1(t_1)\ldots \hat A_N(t_N) \rangle,\qquad t_1\leq \cdots \leq t_N,
\end{equation}
is obtained by recursively applying the QRT. For instance, we define the $n$th auxiliary operator,
\begin{equation}
    \hat\Lambda_n(t_1,\ldots,t_{n-1},t_n,t_n)=\hat\Lambda_{n-1}(t_1,\ldots,t_{n-1},t_n)\hat A_n,
\end{equation}
followed by the auxiliary propagation according to Eq.~\eqref{eq:qrt}. The resulting correlator is given by
\begin{align}
    C(t_1,\ldots,t_N)=\Tr[\hat A_N \hat\Lambda_{N-1}(t_1,\ldots,t_{N-1},t_N)].
\end{align}
The same procedure can be used for an opposite ordering,
\begin{align}
    \langle \hat A_N(t_N)\ldots \hat A_1(t_1) \rangle,
\end{align}
and a symmetric time-ordered correlator,
\begin{equation}
    \langle \hat A_1(t_1)\ldots \hat A_N(t_N)\hat B_N(t_N)\ldots \hat B_1(t_1) \rangle,
\end{equation}
with a slight modification of the auxiliary operator.

\section{\label{sec:3} QRT-GRAPE}
\begin{figure*}
  \centering
  \includegraphics[width=\textwidth]{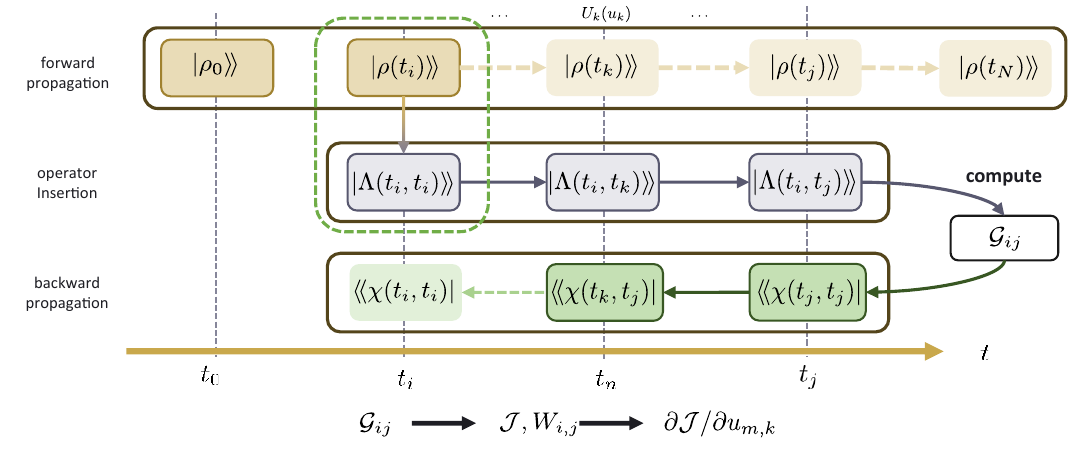}
    \caption{Computation graph for the QRT-GRAPE gradient, illustrated for the auxiliary region case $i<k\le j$. The upper row gives the ordinary forward propagation, the middle row gives the QRT auxiliary-branch initiated by the operator insertion at $t_i$, and the lower row gives propagation of the adjoint variable backward in time from the evaluation of the correlator with $\hat B$ at $t_j$. The derivative of $\mathcal G_{ij}$ with respect to $u_{m,k}$ is obtained by evaluating $\langle\!\langle\chi_{ijk}|(\partial\mathcal U_k/\partial u_{m,k})|\Lambda_{ijk}\rangle\!\rangle$. The definitions of $|\Lambda_{ijk}\rangle\!\rangle$ and $\langle\!\langle\chi_{ijk}|$ for all positions of interval $k$ are given in Eqs.~\eqref{eq:vec_lambda} and~\eqref{eq:vec_chi}, respectively.}
  \label{fig:qrt-grape_computation_graph}
\end{figure*}

We formulate QRT-GRAPE for real-valued objective functions constructed from two-time correlations $\langle\hat A(t_i)\hat B(t_j)\rangle$ on the chronological ordering $i\le j$. Correlators with $i>j$ use the distinct QRT construction in Eq.~\eqref{eq:qrt_opposite_order}. We first vectorize the open-system dynamics and QRT construction, then derive the analytic gradient. This procedure can be easily generalized to multi-time correlators. Figure~\ref{fig:qrt-grape} conceptually compares GRAPE and QRT-GRAPE, and Fig.~\ref{fig:qrt-grape_computation_graph} summarizes the forward, auxiliary, and adjoint propagations used below.

\subsection{\label{sec:3.1} Superoperator Formalism}

Section~\ref{sec:2.2} defines the insertion of the operator and auxiliary evolution. We refer to the auxiliary propagation initiated by this insertion as the \emph{QRT branch}. After vectorization, all procedures in QRT are analogous to those in the state-vector representation. This vectorized representation allows us to combine the QRT with the backward adjoint propagation used in GRAPE.

Given an orthonormal basis $\{\ket{n}\}$, we vectorize an operator $\hat X$ according to
\begin{align}
    \label{eq:convention}
    |n\rangle \langle m| &\mapsto |n\rangle \otimes |m\rangle
\end{align}
\begin{align}
    \label{eq:vectorize}
    \hat X&=\sum_{n,m}X_{nm}\ket{n}\bra{m}\nonumber\\
    &\mapsto|X\rangle\!\rangle := \sum_{n,m} X_{nm}\,\ket{n}\otimes\ket{m}.
\end{align}
Under this representation, left and right multiplication correspond to the following Kronecker products:
\begin{equation}\label{eq:left_right}
  |\hat{A}X\rangle\!\rangle = (\hat{A}\otimes I)|X\rangle\!\rangle, \qquad
  |X\hat{B}\rangle\!\rangle = (I\otimes \hat{B}^T)|X\rangle\!\rangle,
\end{equation}
More generally, $|\hat{A}X\hat{B}\rangle\!\rangle = (\hat{A}\otimes \hat{B}^T)|X\rangle\!\rangle$.
We use this identity for the Lindblad generator and operator insertion below.

Applying the identity to the commutator gives 
\begin{equation}\label{eq:commutator_vec}
  |[ \hat{H},\rho]\rangle\!\rangle = (\hat{H}\otimes I - I\otimes \hat{H}^T)\,|\rho\rangle\!\rangle,
\end{equation}
and the Lindblad equation becomes $\partial_t|\rho\rangle\!\rangle=\mathcal L(t)|\rho\rangle\!\rangle$, where $\mathcal L(t)$ denotes the Liouvillian in this formalism.

The initial condition in QRT $\hat\Lambda(t,t)=\rho(t)\hat A$ is vectorized as
\begin{equation}\label{eq:qrt_init_vec}
  |\Lambda(t,t)\rangle\!\rangle = (I\otimes \hat{A}^T)\,|\rho(t)\rangle\!\rangle.
\end{equation}
By defining the unnormalized maximally entangled vector,
\begin{equation}\label{eq:D_vec}
  |D\rangle\!\rangle := \sum_n \ket{n}\otimes\ket{n},
\end{equation}
the trace takes the inner-product form:
\begin{equation}\label{eq:trace_vec}
  \mathrm{Tr}(X) = \langle\!\langle D|X\rangle\!\rangle, \qquad
  \mathrm{Tr}(\hat{B}X) = \langle\!\langle D|(\hat{B}\otimes I)|X\rangle\!\rangle.
\end{equation}
Therefore, the two-time correlator is
\begin{equation}
    \langle \hat A(t)\hat B(t+\tau) \rangle=\langle\!\langle D|(\hat B\otimes I)|\Lambda(t,t+\tau)\rangle\!\rangle
\end{equation}

\subsection{\label{sec:3.2} Gradient of Two-Time Correlators}

We represent each control function by piecewise constant values $u_{m,k}$ on interval $k$.
The Liouvillian is differentiable with respect to these control values.
We collect the control values in $\bm u$:
\begin{equation}
  \bm u=\{u_{m,k}\}.
\end{equation}
For a single operator pair $\{\hat A,\hat B\}$, we define the two-time correlator for $0\le i\le j\le N$ by
\begin{equation}
    \mathcal{G}_{ij}(\bm u)\coloneq \langle \hat{A}(t_i) \hat{B}(t_j)\rangle
  \label{eq:Gmu_def}
\end{equation}

Using the QRT, we express this correlator in terms of the propagators as
\begin{equation}
  \begin{split}
    \mathcal G_{ij}(\bm u)
    &=
    \langle\!\langle D|\,
    (\hat{B}\otimes I)\,
    \mathcal U_j\cdots \mathcal U_{i+1}
    \\
    &\qquad\times
    (I\otimes \hat{A}^T)\,
    \mathcal U_i\cdots \mathcal U_1\,
    |\rho_0\rangle\!\rangle,
    \quad j\ge i,
  \end{split}
  \label{eq:CAB_discrete}
\end{equation}
where $\mathcal U_k=\exp(\Delta t\mathcal L_k)$ with discretized Lindblad generator $\mathcal L_k$, and $|\rho_0\rangle\!\rangle$ denotes the initial state.
For $i>j$, Eq.~\eqref{eq:qrt_opposite_order} instead uses the earlier-time insertion $\hat B\rho(t_j)$, propagation to $t_i$, and readout with $\hat A$.

Differentiating the correlator with respect to the control value $u_{m,k}$ only affects the sliced propagator $\mathcal U_k$.
With the corresponding auxiliary and adjoint states, the correlator derivative takes the compact form:
\begin{equation}
    \frac{\partial \mathcal G_{ij}(\bm u)}{\partial u_{m,k}}
    =
    \langle\!\langle \chi_{ijk}|
    \frac{\partial \mathcal U_k}{\partial u_{m,k}}
    |\Lambda_{ijk}\rangle\!\rangle,
    \label{eq:CAB_grad_compact}
\end{equation}
where the vectorized operator $|\Lambda_{ijk}\rangle\!\rangle$ is
\begin{equation}\label{eq:vec_lambda}
  |\Lambda_{ijk}\rangle\!\rangle
  :=
  \begin{cases}
    \mathcal U_{k-1}\cdots \mathcal U_1|\rho_0\rangle\!\rangle,
    & k\le i, \\[0.4ex]
    \begin{aligned}[t]
      &\mathcal U_{k-1}\cdots \mathcal U_{i+1}(I\otimes \hat{A}^T)\\
      &\quad\times
      \mathcal U_i\cdots \mathcal U_1|\rho_0\rangle\!\rangle,
    \end{aligned}
    & i<k\le j, \\[0.4ex]
    0,
    & k>j,
  \end{cases}
\end{equation}
and the adjoint variable $\langle\!\langle\chi_{ijk}|$ for the correlator $\mathcal G_{ij}$ is
\begin{equation}\label{eq:vec_chi}
\langle\!\langle \chi_{ijk}|
:=
\begin{cases}
  \begin{aligned}[t]
    &\langle\!\langle D|(\hat{B}\otimes I)\mathcal U_j\cdots \mathcal U_{i+1}\\
    &\quad\times
    (I\otimes \hat{A}^T)\mathcal U_i\cdots \mathcal U_{k+1},
  \end{aligned}
  & k\le i,\\
  \langle\!\langle D|(\hat{B}\otimes I)\mathcal U_j\cdots \mathcal U_{k+1}, & i<k\le j,\\
  0, & k>j.
  \end{cases}
\end{equation}

Using Eq.~\eqref{eq:slice_prop_derivative}, the derivative of the $k$th propagator $\mathcal U_k=e^{\Delta t\mathcal L_k}$ is
\begin{equation}\label{eq:dU_k}
    \frac{\partial \mathcal U_k}{\partial u_{m,k}}
  =\Delta t\int_0^1 e^{(1-s)\Delta t\mathcal L_k}\frac{\partial \mathcal L_k}{\partial u_{m,k}} e^{s\Delta t\mathcal L_k}\,ds.
\end{equation}
For sufficiently small $\Delta t$, the derivative reduces to
\begin{equation}
    \frac{\partial \mathcal U_k}{\partial u_{m,k}}
  \simeq\Delta t\frac{\partial \mathcal L_k}{\partial u_{m,k}}\mathcal U_k.
\end{equation}
In the numerical calculations, we evaluate the exact derivative in Eq.~\eqref{eq:dU_k} using the Fr\'echet derivative of the matrix exponential.
\subsection{Gradient of objective functions constructed from correlations}

In Sec.~\ref{sec:3.2}, we derived the derivative of a single two-time correlator.
To construct objective functions from several correlation functions, we consider the set of two-time correlators $\{\mathcal G^\mu_{ij}(\bm u)\}_{0\le i\le j\le N}$ associated with the operator pairs $\{\hat A_\mu,\hat B_\mu\}$ and define the general objective function by
\begin{equation}
  \mathcal{J}(\bm u)
  \coloneq
  \Phi\left(\{\mathcal G^\mu_{ij}(\bm u)\}\right)
  +
  \mathcal{R}(\bm u),
  \label{eq:J_general}
\end{equation}
where $\Phi$ is a differentiable real-valued function and $\mathcal{R}(\bm u)$ is an optional penalty term. 
The control values $u_{m,k}$ are real and the correlators $\mathcal G_{ij}^{\mu}$ are generally complex, so we use Wirtinger derivatives~\cite{Wirtinger1927Zur,Kreutz-Delgado2009Complex} $\partial_z=\frac{1}{2}(\partial_x-i \partial_y)$ and $\partial_{z^\star}=\frac{1}{2}(\partial_x+i\partial_y)$, defined for $z=x+iy$ by treating $z$ and $z^\star$ as independent variables, which gives
\begin{equation}
  \begin{split}
    \frac{\partial \mathcal{J}}{\partial u_{m,k}}
    &=
    \sum_{\mu}
    \sum_{0\le i\le j\le N}
    \Biggl[
    \frac{\partial \Phi}{\partial \mathcal G_{ij}^\mu(\bm u)}
    \frac{\partial \mathcal G_{ij}^\mu(\bm u)}{\partial u_{m,k}}
    \\
    &\qquad\qquad
    +
    \frac{\partial \Phi}{\partial \mathcal G_{ij}^{\mu *}(\bm u)}
    \frac{\partial \mathcal G_{ij}^{\mu *}(\bm u)}{\partial u_{m,k}}
    \Biggr]
    +
    \frac{\partial \mathcal{R}}{\partial u_{m,k}}.
    \end{split}
    \label{eq:J_chain_complex}
\end{equation}
For real $\Phi$ and real control values, the conjugate terms combine into the real-valued form:
\begin{equation}
  \frac{\partial \mathcal{J}}{\partial u_{m,k}}
  =
  2\,\mathrm{Re}
  \left[
  \sum_{\mu}
  \sum_{0\le i\le j\le N}
  W^{(\mu)}_{ij}\,
  \frac{\partial \mathcal G_{ij}^\mu(\bm u)}{\partial u_{m,k}}
  \right]
  +
  \frac{\partial \mathcal{R}}{\partial u_{m,k}},
  \label{eq:J_chain_real}
\end{equation}
where
\begin{equation}
  W^{(\mu)}_{ij}
  :=
  \frac{\partial \Phi}{\partial \mathcal G_{ij}^\mu(\bm u)}
  \label{eq:sensitivity_weight}
\end{equation}
defines the sensitivity coefficients $W^{(\mu)}_{ij}$.

Combining Eqs.~\eqref{eq:J_chain_real} and \eqref{eq:CAB_grad_compact} gives the QRT-GRAPE gradient,
\begin{equation}
  \boxed{
    \begin{aligned}
      \frac{\partial \mathcal{J}}{\partial u_{m,k}}
      &=
      2\,\mathrm{Re}
      \left[
      \sum_{\mu}
      \sum_{0\le i\le j\le N}
      W^{(\mu)}_{ij}\,
      \langle\!\langle \chi^{(\mu)}_{ijk}|
      \frac{\partial \mathcal U_k}{\partial u_{m,k}}
      |\Lambda^{(\mu)}_{ijk}\rangle\!\rangle
      \right]
      \\
      &\quad+
      \frac{\partial\mathcal R}{\partial u_{m,k}}
    \end{aligned}
  },
  \label{eq:QRT_GRAPE_master_gradient}
\end{equation}
where $|\Lambda^{(\mu)}_{ijk}\rangle\!\rangle$ and $\langle\!\langle \chi^{(\mu)}_{ijk}|$ are the auxiliary and adjoint states associated with the operator pair $\{A_\mu,B_\mu\}$.
The gradient is therefore a linear combination of the derivatives of the two-time correlators, even when $\Phi$ is a nonlinear function.

For objective functions constructed from multi-time correlations, the two-time differentiation procedure can be combined with the recursive multi-time QRT procedure described in Sec.~\ref{sec:2.2} to extend QRT-GRAPE beyond the two-time case.

\section{\label{sec:4} QRT-GRAPE Optimization of Cavity-QED Photon Sources}

In the previous section, we established QRT-GRAPE. In this section, we apply it to heralded remote entanglement generation between two atom--cavity single-photon sources, which has been demonstrated experimentally~\cite{Krutyanskiy2023Entanglement} and analyzed theoretically~\cite{Tanji2024Ratefidelity}. Figure~\ref{fig:cav_qed_fig}(a) shows the atom--cavity system, and Fig.~\ref{fig:cav_qed_fig}(b) shows the entanglement-swapping protocol, in which photons emitted from two such systems interfere at a beam splitter and are detected. Here, we use QRT-GRAPE to optimize the Rabi frequencies driving the two systems.

\subsection{\label{sec:4.1} Atom--Cavity System}

\begin{figure}[tbp]
  \centering
  \includegraphics[width=0.47\textwidth]{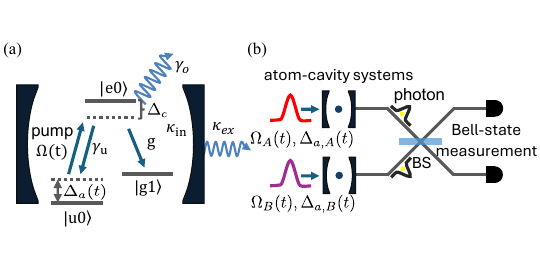}
  \caption{Cavity-QED setting and scope of the photon-source demonstration.
  (a) The numerical model directly propagates each three-level $\Lambda$-type atom--cavity module.
  The pump $\Omega(t)$ drives the $\ket{u0}\leftrightarrow\ket{e0}$ transition, while $\Delta_a(t)$ denotes the controllable diagonal detuning of $\ket{u0}$ in the rotating-frame Hamiltonian.
  The atom--cavity coupling $g$ connects $\ket{e0}$ and $\ket{g1}$, and $\Delta_c$ denotes the cavity detuning.
  The cavity photon leaves through the desired cavity output, characterized by $\kappa_{\mathrm{ex}}$, or is lost internally, characterized by $\kappa_{\mathrm{in}}$.
  Spontaneous emission from $\ket{e0}$ either returns population to $\ket{u0}$, characterized by $\gamma_u$, or transfers it to the zero-excitation state $\ket{g0}$, characterized by $\gamma_o$.
  (b) The broader entanglement-swapping protocol motivates the objective function for temporal mode overlap: photons emitted from two independently driven atom--cavity modules interfere at a beam splitter and are detected in a Bell-state measurement.
  }
  \label{fig:cav_qed_fig}
\end{figure}

 Figure~\ref{fig:cav_qed_fig}(a) shows a three-level $\Lambda$-type atom with ground states $\ket{u}$ and $\ket{g}$ and excited state $\ket{e}$, whose energies are $\hbar\omega_s$ ($s\in\{u,e,g\}$). The transition $\ket{e}\leftrightarrow\ket g$ couples to a single-mode cavity of frequency $\omega_c$ with coupling $g$. An external laser with Rabi frequency $\Omega(t)$ drives the transition $\ket{u}\leftrightarrow\ket{e}$. 

Since the total excitation number is at most one, the coherent dynamics are restricted to $\operatorname{span}\{\ket{u0},\ket{e0},\ket{g1}\}$, where $\ket{s n}=\ket{s}\otimes\ket{n}$ denotes atomic state $\ket{s}\in\{\ket{u},\ket{e},\ket{g}\}$ and cavity Fock state $\ket{n}$. In the basis $\{\ket{u0},\ket{e0},\ket{g1}\}$, the rotating-frame Hamiltonian takes the form
\begin{equation}\label{eq:hamiltonian}
  \hat{H}_{\mathrm{coh}}(t) = \begin{pmatrix}
    \Delta_a(t) & \Omega(t) & 0\\
    \Omega(t) & 0 & g\\
    0 & g & \Delta_c
  \end{pmatrix},
\end{equation}
where $\Delta_c=\omega_c-\omega_e+\omega_g$ denotes a cavity detuning, and $\Delta_a(t) = \omega_L(t)-\omega_e+\omega_u$ denotes an atomic detuning with instantaneous laser frequency $\omega_L(t)$.

Spontaneous decay rate from $\ket{e0}$ back to $\ket{u0}$ and to $\ket{g0}$ are characterized by $\gamma_u$ and $\gamma_o$, respectively, while cavity decay from $\ket{g1}$ is characterized by $\kappa_{\mathrm{ex}}$ and $\kappa_{\mathrm{in}}$ for cavity output and internal cavity loss.
To preserve trace, we include the zero-excitation state $\ket{g0}$ to the Hilbert space:
\begin{equation}
  \mathcal H
  =
  \operatorname{span}
  \left\{
  \ket{u0},
  \ket{e0},
  \ket{g1},
  \ket{g0}
  \right\}.
\end{equation}
Here, $\ket{g0}$ denotes the zero-excitation state reached after cavity decay or spontaneous emission by $\gamma_o$. On this four-state space, $\hat H(t)=\hat H_{\mathrm{coh}}(t)\oplus0_{\ket{g0}}$, and the dynamics are governed by the Lindblad master equation,
\begin{align}\label{eq:example_master}
  \dot{\rho}(t)
  &=
  -i[\hat{H}(t),\rho(t)]
  +
  \sum_{\nu}
  \mathcal D[\hat{L}_\nu]\rho(t),
  \\
  \mathcal D[\hat{L}]\rho
  &=
  \hat{L}\rho \hat{L}^\dagger
  -\frac{1}{2}\{\hat{L}^\dagger \hat{L},\rho\},
  \nonumber
\end{align}
where the index $\nu\in\{u,o,\mathrm{ex},\mathrm{in}\}$ labels the Lindblad operators defined below:
\begin{align}
  \hat{L}_u
  &=
  \sqrt{2\gamma_u}\ket{u0}\!\bra{e0},
  &
  \hat{L}_o
  &=
  \sqrt{2\gamma_o}\ket{g0}\!\bra{e0},
  \nonumber\\
  \hat{L}_{\mathrm{ex}}
  &=
  \sqrt{2\kappa_{\mathrm{ex}}}\ket{g0}\!\bra{g1},
  &
  \hat{L}_{\mathrm{in}}
  &=
  \sqrt{2\kappa_{\mathrm{in}}}\ket{g0}\!\bra{g1}.
\end{align}

Here, $\gamma\equiv\gamma_u+\gamma_o$ is the amplitude decay rate of $\ket{e0}$ due to spontaneous emission, so the corresponding population decay rate is $2\gamma$.
Hereafter, for the cavity-QED example and its numerical appendices, we express frequencies (including detunings), coupling strengths, decay rates, and Rabi frequencies in units of $\gamma$, and times in units of $\gamma^{-1}$.
We use the same symbols for the resulting dimensionless quantities.

The $\hat{L}_u$ returns population to $\ket{u0}$, allowing subsequent re-excitation, whereas $\hat{L}_o$ transfers population to $\ket{g0}$ without contributing to the desired cavity output. Both cavity output and internal cavity loss transfer population from $\ket{g1}$ to $\ket{g0}$, with only $\hat{L}_{\mathrm{ex}}$ counted as useful output.

\subsection{\label{sec:4.2} Objective functions for remote entanglement generation}
We apply QRT-GRAPE to atom--photon entanglement generation for cavity-based entanglement swapping, as illustrated in Fig.~\ref{fig:cav_qed_fig}(b).
The emitted photons interfere at the central node and are subsequently measured by single-photon detectors. The success probability and fidelity of atom-atom entanglement are determined by the generation probability of atom--photon entanglement and the first-order coherence of each photon, respectively. Therefore, the numerical model treats the two independent atom--cavity sources and evaluates their output-field correlations and photon-emission probabilities. The noises outside the atom-cavity systems are neglected since they are not relevant to the optimization problem.

We consider two atom--cavity systems A and B. For the system $\mu\in\{A,B\}$, the photon-emission probability $P_{\mathrm{ex},\mu}$ is obtained from the output photon flux as
\begin{equation}\label{eq:emission probability}
  P_{\mathrm{ex},\mu}
  =
  2\kappa_{\mathrm{ex},\mu}
  \int_0^\infty dt\,
  \langle \hat{a}^\dagger(t)\hat{a}(t)\rangle_\mu .
\end{equation}
We quantify the temporal mode indistinguishability of the two emitted photons by the normalized wave-packet overlap $J_{AB}$: 
\begin{equation}\label{eq:indistinguishability}
  \begin{aligned}
    J_{AB}
    &=\frac{4\kappa_{\mathrm{ex},A}\kappa_{\mathrm{ex},B}}{P_{\mathrm{ex},A}P_{\mathrm{ex},B}} \\
    & \quad \times
    \int_0^\infty\!\!\int_0^\infty dt\,dt'\; \langle \hat{a}^\dagger(t)\hat{a}(t')\rangle_A \langle \hat{a}^\dagger(t')\hat{a}(t)\rangle_B,
  \end{aligned}
\end{equation}
When the  temporal mode mismatch is the only noise source, $J_{AB}$ determines the conditional Bell-state fidelity~\cite{Tanji2024Ratefidelity}:
\begin{equation}\label{eq:conditional_entanglement_fidelity}
  F_{\mathrm{Bell}}
  =
  \frac{1}{2}\mathrm{Re}\left[1+J_{AB}\right].
\end{equation}
Maximizing the real part of $J_{AB}$ improves the conditional fidelity but does not reward photon-generation probability.
To combine photon generation with  temporal mode overlap in a single optimization target, we use the unnormalized overlap~\cite{Tanji2024Ratefidelity}:
\begin{equation}\label{eq:objective}
  \begin{aligned}
    \mathcal{J}
    &=
    P_{\mathrm{ex},A}P_{\mathrm{ex},B}J_{AB} \\
    &=
    4\kappa_{\mathrm{ex},A}\kappa_{\mathrm{ex},B}
    \int\!\!\int_0^\infty dt\,dt'\;
    \langle \hat{a}^\dagger(t)\hat{a}(t')\rangle_A
    \langle \hat{a}^\dagger(t')\hat{a}(t)\rangle_B .
  \end{aligned}
\end{equation}
The prefactors follow from the input-output relation $\hat{a}_{\mathrm{out}}(t)=\sqrt{2\kappa_{\mathrm{ex}}}\,\hat{a}(t)$.
For each source,
\begin{equation}
    G^{(\mu)}(t,t')=\langle\hat{a}^\dagger(t)\hat{a}(t')\rangle_\mu=G^{(\mu)*}(t',t),
\end{equation}
so exchanging $t$ and $t'$ maps the correlation to its complex conjugate and makes $\mathcal J$ real.

Hereafter, we discretize the objective function for numerical optimization.
Let $t_i$ denote the $i$th time point at which the two-time correlation functions are evaluated, and let $w_i$ denote the trapezoidal integration weight at $t_i$. The discretized objective function is given by Eq.~\eqref{eq:objective_discrete}.
\begin{equation}\label{eq:objective_discrete}
  \mathcal{J}
  =
  4\kappa_{\mathrm{ex},A}\kappa_{\mathrm{ex},B}
  \sum_{i,j=0}^{M-1}
  w_iw_j
  \langle \hat{a}^\dagger(t_i)\hat{a}(t_j)\rangle_A
  \langle \hat{a}^\dagger(t_j)\hat{a}(t_i)\rangle_B.
\end{equation}
Writing $G^{(\mu)}_{ij}=\langle \hat a^\dagger(t_i)\hat a(t_j)\rangle_\mu$, we use the Hermitian symmetry $G^{(\mu)}_{ji}=G^{(\mu)*}_{ij}$ to obtain the upper-triangular form:
\begin{equation}\label{eq:objective_discrete_upper}
  \begin{aligned}
    \mathcal J
    ={}&4\kappa_{\mathrm{ex},A}\kappa_{\mathrm{ex},B}
    \Biggl[
    \sum_{i=0}^{M-1} w_i^2 G^{(A)}_{ii}G^{(B)}_{ii}
    \\
    &\quad+2\,\mathrm{Re}
    \sum_{0\le i<j\le M-1}
    w_iw_j G^{(A)}_{ij}G^{(B)*}_{ij}
    \Biggr].
  \end{aligned}
\end{equation}
The trapezoidal weights are given by
\begin{equation}\label{eq:objective_quadrature_weights}
  w_i
  =
  \begin{cases}
    (t_1-t_0)/2, & i=0,\\
    (t_{i+1}-t_{i-1})/2, & 1\le i\le M-2,\\
    (t_{M-1}-t_{M-2})/2, & i=M-1.
  \end{cases}
\end{equation}
The objective function $\mathcal J$ is evaluated using the correlation functions at the $M$ time points $t_0,\ldots,t_{M-1}$. The subsequent time $t_M$ is the final edge of the last simulation interval and is not itself a correlation calculation point.

For source $\mu\in\{
A,B\}$, we decompose the $k$th-interval Hamiltonian into the static term and two control Hamiltonians:
\begin{align}\label{eq:real_control_example}
\hat{H}_k^{(\mu)}
=
\hat{H}_0^{(\mu)}
+\sum_{m=1}^2 u^{(\mu)}_{m,k}\hat{H}_m^{(\mu)},
\end{align}
where $u^{(\mu)}_{1,k}$ and $u^{(\mu)}_{2,k}$ are the constant values of the Rabi frequency $\Omega_\mu(t)$ and atomic detuning $\Delta_{a}(t)$ on $[t_{k-1},t_k)$.
The corresponding control Hamiltonians are
\begin{equation}
  \hat{H}_1^{(\mu)}
  =
  \ket{u0}\!\bra{e0}+\ket{e0}\!\bra{u0},
  \qquad
  \hat{H}_2^{(\mu)}
  =
  \ket{u0}\!\bra{u0},
\end{equation}
while $\hat{H}_0^{(\mu)}$ contains the cavity detuning and atom--cavity coupling.

Because the two sources evolve independently, differentiating $\mathcal J$ with respect to $u^{(A)}_{m,k}$ acts only on the correlator for the source A. Then, the correlator for the source B, $G^{(B)*}_{ij}$, is treated as a scalar weight, and vice versa.

Since the static Hamiltonian is independent of the control values, the derivative of the Liouvillian with respect to either control value is determined solely by the corresponding control Hamiltonian:
\begin{equation}\label{eq:control_liouvillian_direction}
  \frac{\partial\mathcal L_k^{(\mu)}}{\partial u^{(\mu)}_{m,k}}
  =
  -i\left[
  \hat H_m^{(\mu)}\otimes\mathbb I
  -
  \mathbb I\otimes\hat H_m^{(\mu)T}
  \right].
\end{equation}
We evaluate the propagator derivative using Eq.~\eqref{eq:dU_k}, with the Liouvillian derivative in Eq.~\eqref{eq:control_liouvillian_direction}.
For source A, Eq.~\eqref{eq:CAB_grad_compact} gives
\begin{equation}\label{eq:dcorr}
\frac{\partial G^{(A)}_{ij}}{\partial u^{(A)}_{m,k}}
=\langle\!\langle\chi^{(A)}_{ijk}|\frac{\partial\mathcal U^{(A)}_k}{\partial u^{(A)}_{m,k}}|\Lambda^{(A)}_{ijk}\rangle\!\rangle.
\end{equation}
Here $|\Lambda^{(A)}_{ijk}\rangle\!\rangle$ denotes the vectorized auxiliary operator at interval $k$, while $\langle\!\langle\chi^{(A)}_{ijk}|$ denotes the adjoint state for the correlator $G^{(A)}_{ij}$, as given in Eqs.~\eqref{eq:vec_lambda} and~\eqref{eq:vec_chi}, respectively.
For notational simplicity, we denote this correlator derivative by
\begin{equation}
D^{(A)}_{ij,mk}:=\frac{\partial G^{(A)}_{ij}}{\partial u^{(A)}_{m,k}}.
\end{equation}
Note that $D^{(A)}_{ij,mk}=0$ for $k>j$ since the correlator is independent of the control after $t_j$.

For source $A$, the off-diagonal terms in Eq.~\eqref{eq:objective_discrete_upper} give $W^{(A)}_{ij}=4\kappa_{\mathrm{ex},A}\kappa_{\mathrm{ex},B}w_iw_jG^{(B)*}_{ij}$ for $i<j$.
For the real diagonal correlators in this application, we extend the objective to depend on $G^{(\mu)}_{ii}$ through $\mathrm{Re}\,G^{(\mu)}_{ii}$, giving $W^{(A)}_{ii}=2\kappa_{\mathrm{ex},A}\kappa_{\mathrm{ex},B}w_i^2G^{(B)}_{ii}$.
Substitution into Eq.~\eqref{eq:J_chain_real} gives the gradient with respect to $u^{(A)}_{m,k}$:
\begin{align}\label{eq:full_gradient}
  \frac{\partial \mathcal{J}}{\partial u^{(A)}_{m,k}}
  &=
  8\kappa_{\mathrm{ex},A}\kappa_{\mathrm{ex},B}
  \sum_{\substack{i<j\\j\ge k}}
  w_iw_j
  \mathrm{Re}\!\left[
  G^{(B)*}_{ij}\,D_{ij,mk}^{(A)}
  \right] \nonumber\\
  &\quad +
  4\kappa_{\mathrm{ex},A}\kappa_{\mathrm{ex},B}
  \sum_{i\ge k}
  w_i^2
  \mathrm{Re}\!\left[
  G^{(B)*}_{ii}\,D_{ii,mk}^{(A)}
  \right].
\end{align}
The same expression applies to source B by changing $A$ and $B$. 

\subsection{\label{sec:optimization_results}Optimization results of entanglement swapping}

\begin{table}[tbp]
  \centering
  \caption{Physical parameters used for the two atom--cavity sources in the numerical demonstration. All frequencies and decay rates are normalized by $\gamma\equiv\gamma_u+\gamma_o$.}
  \label{tab:physical_params}
  \small
  \begin{tabular}{@{}lcc@{}}
    \toprule
    Parameter & Cavity A & Cavity B \\
    \midrule
    $g$ & 1.4 & 1.0 \\
    $\Delta_c$ & 0 & 0 \\
    $\kappa_{\mathrm{ex}}$ & 0.9 & 1.1 \\
    $\kappa_{\mathrm{in}}$ & 0.08 & 0.08 \\
    $\gamma_u$ & 0.3 & 0.3 \\
    $\gamma_o$ & 0.7 & 0.7 \\
    \bottomrule
  \end{tabular}
\end{table}

The parameters used in the examples are listed in Table~\ref{tab:physical_params}. For the comparison in Table~\ref{tab:main-score-comparison}, we optimize the Rabi frequencies $\Omega_A(t)$ and $\Omega_B(t)$ with both atomic detunings fixed at zero.

We divide the Rabi pulse driving period $0\leq t<10$ for sources A and B into $100$ equal time intervals, with one constant Rabi frequency for each source in each interval. The Rabi frequency of each source in the first interval is fixed at zero. QRT-GRAPE optimizes the Rabi frequencies in the remaining $99$ intervals of each source, giving $198$ adjustable Rabi frequencies in total. Both atomic detunings remain fixed at zero.
Each adjustable Rabi frequency is bounded by $[-2,2]$.
Both Rabi frequencies and both atomic detunings are zero during an additional decay interval, allowing the remaining cavity photon to decay. We use box-constrained L-BFGS-B with the QRT-GRAPE gradient. Appendix~\ref{app:numerical_protocol} gives the further numerical settings.

We considered Gaussian and flat-top Rabi pulses for comparison to QRT-GRAPE. For the Gaussian Rabi pulses, we vary the pulse amplitudes and widths. For the flat-top Rabi pulses, we vary the pulse amplitudes and durations (see details in Appendix~\ref{app:numerical_protocol}).
For each pulse shape, we select the A/B pulse pair with the highest objective function value $\mathcal J$ among the candidate pairs. Starting from each selected pulse pair, QRT-GRAPE optimizes $99$ Rabi frequencies per source, with both atomic detunings fixed at zero.
Before QRT-GRAPE optimization, the selected pairs of Gaussian and flat-top Rabi pulses gave objective function values $\mathcal J=0.2354$ and $0.2441$, respectively. Starting from either selected pulse pair, QRT-GRAPE reached $\mathcal J=0.2857$, corresponding to relative increases of $21.4\%$ and $17.0\%$ for the Gaussian and flat-top pulses, respectively.
Table~\ref{tab:main-score-comparison} summarizes the corresponding objective function values $\mathcal J$, conditional Bell-state fidelity, and external photon-emission probabilities before and after optimization.

\begin{table}[t]
  \centering
  \small
  \setlength{\tabcolsep}{3.5pt}
  \caption{Comparison of the objective function value $\mathcal{J}$, conditional Bell-state fidelity $F_\mathrm{Bell}$, and external photon-emission probabilities $P_{\mathrm{ex,A}}$ and $P_{\mathrm{ex,B}}$ before and after QRT-GRAPE optimization, for the physical parameters in Table~\ref{tab:physical_params}. We optimize the discretized Rabi frequencies $\Omega_A(t)$ and $\Omega_B(t)$ while fixing both atomic detunings at zero. Fixing the first Rabi frequency of each source at zero leaves $198$ adjustable Rabi frequencies.}
  \label{tab:main-score-comparison}
  \begin{tabular}{@{}llrrrr@{}}
  \toprule
  Initialization & Stage & $\mathcal{J}$ & $F_{\mathrm{Bell}}$ & $P_{\mathrm{ex},A}$ & $P_{\mathrm{ex},B}$ \\
  \midrule
  Gaussian & Initial & 0.2354 & 0.9401 & 0.5957 & 0.4489 \\
           & Optimized & 0.2857 & 0.9441 & 0.6569 & 0.4896 \\
  Flat-top & Initial & 0.2441 & 0.9356 & 0.6065 & 0.4620 \\
           & Optimized & 0.2857 & 0.9441 & 0.6569 & 0.4896 \\
  \bottomrule
  \end{tabular}
\end{table}

Appendix~\ref{app:optimization_behavior} shows how the objective function value $\mathcal J$ and the norm of its gradient with respect to the adjustable Rabi frequency change during optimization starting from the selected Gaussian and flat-top Rabi pulse pairs. We also performed 100 QRT-GRAPE optimizations starting from independently generated random Rabi pulses, with both atomic detunings fixed at zero. 
These optimizations are described in Appendix~\ref{app:random_rabi}. Of the 100 optimizations, 87 ended with a final objective function value of $\mathcal J\geq0.285$, showing that high objective function values were reached from many different initial Rabi pulse pairs. Some optimizations terminated at lower values. These results suggest that running the optimization from several initial Rabi pulse pairs and retaining the largest final value of $\mathcal J$ can help reduce dependence on initialization in this example.

We also tested whether penalizing differences between neighboring Rabi frequencies reduces steep variations in the optimized pulses. Starting from the Gaussian Rabi pulse pair selected for Table~\ref{tab:main-score-comparison}, we optimized only the Rabi frequencies, with both atomic detunings fixed at zero (Appendix~\ref{app:smoothness_regularization}). Penalizing differences between neighboring Rabi frequency values produced smoother optimized pulses, with an approximately $0.24\%$ decrease in $\mathcal J$ for $\lambda=10^{-3}$ relative to optimization without the penalty. In this example, QRT-GRAPE therefore found smoother pulses while retaining nearly the same objective function value. The roughness measure $R[u]$ used in this test excludes the change in each Rabi frequency when the drive is turned off. We expect that including this change in the penalty would also reduce the magnitude of the jump at the end of the drive.

So far, we have considered the nonidentical atom--cavity systems with the fixed parameters in Table~\ref{tab:physical_params}. In practical networks, the parameters of the two atom--cavity systems can differ~\cite{Kikura2026Passive,Nolleke2013Efficient}. We therefore investigate how our method improves the objective function value for various combinations of $g_B/g_A$ and $\kappa_{\text{ex},B}/\kappa_{\text{ex},A}^{\text{opt}}$. In the simulation, we use the parameters of system A in Table~\ref{tab:physical_params} except for $\kappa_{\mathrm{ex},A}$, while the $g_B$ and $\kappa_{\text{ex},B}$ are swept such that $g_B/g_A,\kappa_{\text{ex},B}/\kappa_{\text{ex},A}^{\text{opt}}\in [0.5,1.0]$. 
Since the photon-emission probability depends on the cavity decay rate $\kappa_\text{ex}$, we choose the optimal cavity-decay rate for the cavity A, which maximizes the photon-emission probability~\cite{Goto2019Figure}:
\begin{equation}
\kappa_{\mathrm{ex},A}^{\mathrm{opt}}
=
\kappa_{\mathrm{in},A}
\sqrt{
1+\frac{g_A^2}{\kappa_{\mathrm{in},A}\gamma_{o,A}}
}.
\end{equation}
Substituting the parameters given in Table~\ref{tab:physical_params}, the optimal decay rate becomes $0.48$.

\begin{figure*}[tbp]
  \centering
  \includegraphics[width=\textwidth]{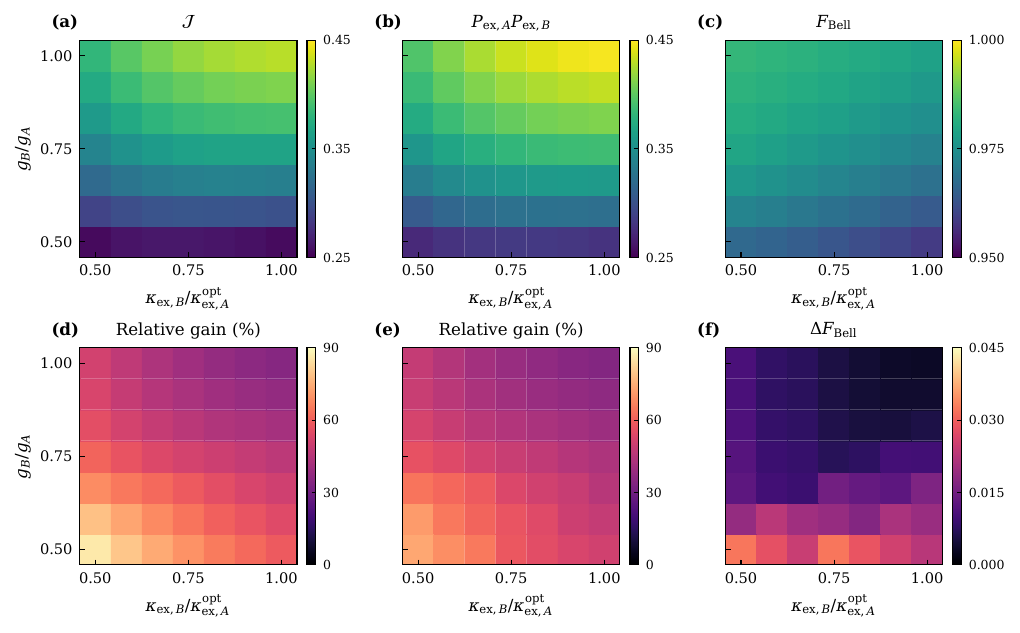}
  \caption{Objective function value, product of photon-emission probabilities, and Bell-state fidelity after QRT-GRAPE optimization as a function of the parameters of source B.
  The horizontal and vertical axes are $\kappa_{\mathrm{ex},B}/\kappa_{\mathrm{ex},A}^{\mathrm{opt}}$ and $g_B/g_A$, respectively, with $g_A=1.4$ and $\kappa_{\mathrm{ex},A}^{\mathrm{opt}}=0.48$.
  %At each of the $49$ grid points, a sweep over Gaussian pulse amplitudes and widths selects the pair of Rabi pulses with the highest achieved value of $\mathcal J$.
  %QRT-GRAPE starts once from that selected pair and optimizes the Rabi pulses with both atomic detunings fixed at zero.
  (a--c) Optimized objective function value $\mathcal J$, photon-emission probability product $P\equiv P_{\mathrm{ex},A}P_{\mathrm{ex},B}$, and conditional Bell-state fidelity $F_{\mathrm{Bell}}$, respectively.
  (d,e) Relative gains $100(X_{\mathrm{opt}}-X_{\mathrm G})/X_{\mathrm G}$ in percent for $X=\mathcal J$ and $P$, respectively.
  (f) Dimensionless absolute difference $\Delta F_{\mathrm{Bell}}=F_{\mathrm{Bell,opt}}-F_{\mathrm{Bell,G}}$.
  The subscript $\mathrm G$ denotes the corresponding value for the Gaussian Rabi pulse pair selected at the same grid point.
  All three quantities increase at every grid point relative to their values for that Gaussian Rabi pulse pair.}
  \label{fig:result_heatmap}
\end{figure*}

In Fig.~\ref{fig:result_heatmap}, the atom--cavity coupling ratio $g_B/g_A$ and external cavity decay rate ratio $\kappa_{\mathrm{ex},B}/\kappa_{\mathrm{ex},A}^{\mathrm{opt}}$ each take seven equally spaced values from $0.5$ to $1.0$, giving a $7\times7$ grid. At each grid point, we construct Gaussian Rabi pulses for each source from $20$ amplitudes in $[0.1,2]$ and $20$ widths in $[0.1,T/6]$, with $T=10$ and pulse center $T/2$. Appendix~\ref{app:numerical_protocol} defines the Gaussian Rabi pulses and describes how they are discretized into 100 piecewise-constant time intervals. Each source has $400$ candidate pulses; we evaluate the objective function $\mathcal J$ for all $400^2$ pairs of Gaussian pulses from sources A and B. The pair with the highest value of $\mathcal J$ is denoted by $\mathcal J_{\mathrm G}$. At each grid point, we run QRT-GRAPE once from that selected pair, optimizing the Rabi pulses while keeping both atomic detunings fixed at zero.

Figure~\ref{fig:result_heatmap}(a) shows the optimized objective function value $\mathcal J_{\mathrm{opt}}$.
Figures~\ref{fig:result_heatmap}(b) and \ref{fig:result_heatmap}(c) show the corresponding optimized photon-emission probability product $P\equiv P_{\mathrm{ex},A}P_{\mathrm{ex},B}$ and conditional Bell-state fidelity $F_{\mathrm{Bell}}$, respectively.
The upper-right grid point $(1,1)$ corresponds to identical atom--cavity sources.
With $\kappa_{\mathrm{ex},B}/\kappa_{\mathrm{ex},A}^{\mathrm{opt}}=1$, reducing $g_B/g_A$ from $1$ to $0.5$ decreases $\mathcal J_{\mathrm{opt}}$ by $40.2\%$.
With $g_B/g_A=1$, reducing $\kappa_{\mathrm{ex},B}/\kappa_{\mathrm{ex},A}^{\mathrm{opt}}$ from $1$ to $0.5$ decreases $\mathcal J_{\mathrm{opt}}$ by $10.9\%$.
% At weaker $g_B$, panel (a) shows the dependence on $\kappa_{\mathrm{ex},B}$ is nonmonotonic, showing that the two varied parameters do not follow a simple separable trend over the sampled grid.

Figures~\ref{fig:result_heatmap}(d) and \ref{fig:result_heatmap}(e) show the relative gains in the objective function value $\mathcal J$ and photon-emission probability product $P\equiv P_{\mathrm{ex},A}P_{\mathrm{ex},B}$, respectively.
The relative gain in $\mathcal J$ ranges from $34.8\%$ to $86.0\%$, with a median of $52.5\%$.
The relative gain in $P$ ranges from $33.8\%$ to $73.2\%$, with a median of $48.9\%$.
Figure~\ref{fig:result_heatmap}(f) shows the absolute Bell-state fidelity difference $\Delta F_{\mathrm{Bell}}$, which ranges from $0.0032$ to $0.0322$, with a median of $0.0098$.
All three increases are positive at all $49$ grid points.
Every comparison uses the Rabi pulses optimized at a particular grid point and the Gaussian Rabi pulses selected at that same grid point.

Reducing the atom--cavity coupling lowers the internal cooperativity and the upper bound on the single-photon generation efficiency~\cite{Goto2019Figure}.
This is consistent with the substantial decrease in the objective function value at weaker coupling even after optimization of the Rabi pulses.
The external cavity decay rate mismatch produces a smaller decrease in the optimized objective function value than the reduction in coupling strength.
Over the parameter range studied here, these results suggest that weak atom--cavity coupling is the main factor limiting the optimized performance, rather than the external cavity decay rate mismatch.

We also compared optimization of the Rabi frequencies alone with joint optimization of the Rabi frequencies and atomic detunings. Starting from the Gaussian Rabi pulse pair selected for Table~\ref{tab:main-score-comparison}, we performed $30$ joint optimizations with independently generated initial atomic detuning values (Appendix~\ref{app:detuning_initialization}). None of the 30 optimizations of both Rabi frequencies and atomic detunings exceeded the existing result obtained by optimizing the Rabi frequencies alone.

We verified the analytic gradient by comparing it with centered finite differences with respect to selected adjustable values of the Rabi frequencies and atomic detunings of both sources.
The analytic derivatives of $\mathcal J$ agreed with the centered-finite-difference estimates of the corresponding partial derivatives of $\mathcal J$ to high numerical accuracy, supporting the derivative construction and its numerical implementation.
In the timing comparison, evaluating $\mathcal J$ and its analytic gradient was 9.40 times as fast as computing the full gradient by centered finite differences.
Further details of the validation are provided in Appendix~\ref{app:gradient_verification}.

\section{\label{sec:5} Conclusion}

In this paper, we developed QRT-GRAPE for optimizing objective functions constructed from multi-time correlation functions.
The method incorporates the operator insertion and auxiliary propagation in QRT into the forward propagation and backward adjoint propagation used in GRAPE.
This construction provides analytic gradients for objective functions defined directly by multi-time correlation functions, extending gradient-based optimal control beyond objective functions represented by the state at a single final time.

As an application of QRT-GRAPE, we considered heralded remote entanglement generation using two nonidentical atom--cavity single-photon sources.
Starting from the flat-top Rabi pulses, QRT-GRAPE increased the objective function value $\mathcal J$ from $0.2441$ to $0.2857$, corresponding to a relative improvement of $17.0\%$. In the same comparison, the photon-emission probability product $P_{\mathrm{ex},A}P_{\mathrm{ex},B}$ increased from $0.2802$ to $0.3216$, while the Bell-state fidelity $F_{\mathrm{Bell}}$ increased from $0.9356$ to $0.9441$.

We also varied the atom--cavity coupling ratio $g_B/g_A$ and external cavity decay rate ratio $\kappa_{\mathrm{ex},B}/\kappa_{\mathrm{ex},A}^{\mathrm{opt}}$ over a $7\times7$ grid.
At each of the $49$ grid points, we selected the Gaussian Rabi pulse pair with the highest value of $\mathcal J$ from the Gaussian pulse amplitude and width search and ran one QRT-GRAPE optimization starting from that pair. At every grid point, QRT-GRAPE increased the objective function value $\mathcal J$, photon-emission probability product, and Bell-state fidelity relative to the values obtained with the Gaussian Rabi pulse pair selected at the same grid point. The relative increase in $\mathcal J$ ranged from $34.8\%$ to $86.0\%$.

The present formulation assumes that, after each operator insertion required by the QRT construction, the auxiliary operator evolves under the same time-dependent generator that governs the density operator.
Extending QRT-GRAPE to cases in which this QRT propagation rule is not valid will require a different treatment of multi-time correlations.
A further practical challenge is the computational cost of evaluating correlations over many combinations of times, motivating more efficient propagation and selection of time points for evaluating multi-time correlation functions.
Subject to this QRT condition, QRT-GRAPE provides a general framework for optimizing objective functions constructed from multi-time correlation functions.

\section*{Data Availability}

The data that support the findings of this study are available from the corresponding author upon reasonable request.

\begin{acknowledgments}
This work was supported by JST Moonshot R\&
D, Grant
Number JPMJMS256E and JPMJMS256G, JST ASPIRE, Grant Number JPMJAP2427, JST COI--NEXT Grant No. JPMJPF2221, and Program for the Advancement of Next Generation Research Projects, Keio University. K.T. was supported by JST SPRING, Grant Number JPMJSP2123.
\end{acknowledgments}

\appendix
\section{\label{app:numerical_protocol} Optimization and initialization methods}
In this Appendix, we describe the optimization settings and methods. 
Table~\ref{tab:numerical_protocol} summarizes the settings for Table~\ref{tab:main-score-comparison} and Fig.~\ref{fig:result_heatmap}. We divide $0\leq t<T=10$ into $100$ intervals of width $\Delta t=0.1$. For each Rabi frequency and atomic detuning, we use one value at the left endpoint of each interval and hold that value constant throughout the interval. The value of each function in the first time interval is fixed at zero. Both Rabi frequencies and both atomic detunings are zero during the additional decay interval from $10$ to approximately $14.2373$, which is divided into $10$ intervals.

For Table~\ref{tab:main-score-comparison} and Fig.~\ref{fig:result_heatmap}, we optimize the remaining $99$ values of each Rabi frequency, $\Omega_A(t)$ and $\Omega_B(t)$, within $[-2,2]$, giving $198$ adjustable Rabi frequencies, while both atomic detunings remain fixed at zero. The optimizations from random initial Rabi pulses described below and the smoothness test each optimize $99$ Rabi frequencies per source, with both atomic detunings fixed at zero, giving $198$ adjustable values per optimization. For the detuning comparison in Appendix~\ref{app:detuning_initialization}, we reuse the result obtained by optimizing the Rabi frequencies with both atomic detunings fixed at zero, and perform $30$ new optimizations with $396$ adjustable control values each.

For Table~\ref{tab:main-score-comparison}, we first select pairs of Gaussian Rabi pulses and pairs of flat-top Rabi pulses by varying their amplitudes and Gaussian widths or flat-top durations with both atomic detunings fixed at zero, as described below. QRT-GRAPE then allows the Rabi frequency in each of the remaining $99$ time intervals of each photon source to vary. The optimized pulses can therefore depart from Gaussian or flat-top shapes. We use box-constrained L-BFGS-B with the analytic QRT-GRAPE gradient of $\mathcal J$ and at most $500$ iterations. 

\begin{table}[tbp]
  \centering
  \small
  \caption{Numerical settings used for the two-source optimization and the cavity parameter study as shown in Table~\ref{tab:main-score-comparison}. The Gaussian width range is used to select the initial Gaussian pulses. All time quantities are expressed in units of $1/\gamma$, where $\gamma=\gamma_u+\gamma_o$.}
  \begin{tabular}{@{}ll@{}}
    \toprule
    Parameter & Value \\
    \midrule
    Rabi-frequency bounds & $[-2,2]$ \\
    Atomic detunings & Fixed at zero \\
    \shortstack[l]{Number of adjustable\\Rabi frequencies} & $198$ \\
    Initial Gaussian width $\sigma$ & $[0.1,10/6]$ \\
    \shortstack[l]{Duration of the Rabi\\pulse driving period} & $10$ \\
    Additional decay duration & $\approx 4.2373$ \\
    \shortstack[l]{Number of time intervals\\in the Rabi pulse driving period} & $100$ \\
    Decay intervals & $10$ \\
    Maximum iterations & $500$ \\
    \bottomrule
  \end{tabular}
  \label{tab:numerical_protocol}
\end{table}

\subsection{Initial Rabi Pulse Pairs}
We selected the initial Gaussian and flat-top Rabi pulse pairs for Table~\ref{tab:main-score-comparison} as follows.
The Rabi pulses were fixed at zero at $t=0$ and for $t\geq T=10$.
Both atomic detunings, $\Delta_{a,A}(t)$ and $\Delta_{a,B}(t)$, were fixed at zero during the pulse parameter sweeps and the QRT-GRAPE optimizations reported in Table~\ref{tab:main-score-comparison}.

For Table~\ref{tab:main-score-comparison} and Fig.~\ref{fig:result_heatmap}, the initial Gaussian Rabi pulse for source $\mu\in\{A,B\}$ was obtained by discretizing the function
\begin{equation}
  \Omega_{\mu}^{\mathrm G}(t)=
  \begin{cases}
    A_{\mu}\exp\!\left[-\dfrac{(t-T/2)^2}{2\sigma_{\mu}^2}\right],
      & 0<t<T,\\
    0, & t=0\ \text{or}\ t\geq T,
  \end{cases}
  \label{eq:gaussian_initializer}
\end{equation}
with $T=10$, center $T/2$, and $\sigma_{\mu}\in[0.1,T/6]$. We evaluated $\Omega_\mu^{\mathrm G}(t)$ at the left endpoint of each time interval and held that Rabi frequency constant over the interval. Both Rabi frequencies and both atomic detunings were zero during the additional decay interval.

The flat-top pulse $\Omega_\mu^F(t)$ is defined by
\begin{equation}
    \frac{\Omega_\mu^F(t)}{A_{\mu}}=\begin{cases}
        \sin^2\left[\frac{\pi t}{2r\tau_{\mu}}\right] & 0\leq t<r\tau_{\mu},\\
        1 &r\tau_{\mu}\leq t\leq(1-r)\tau_{\mu},\\
        \sin^2\left[\pi\frac{\tau_{\mu}-t}{2r\tau_{\mu}}\right] &(1-r)\tau_{\mu}<t\leq \tau_{\mu},
    \end{cases}
\end{equation}
where $r=0.1$ and $\tau_\mu=D_\mu$ is the total flat-top pulse duration, including the rise and fall.
The pulse is zero outside $[0,\tau_\mu]$.

For each source in Table~\ref{tab:main-score-comparison}, we constructed Gaussian Rabi pulses using $100$ uniformly spaced amplitudes in $[0.1,2]$ and $100$ uniformly spaced widths in $[0.1,T/6]$. For the flat-top Rabi pulses, we used the $100$ Rabi frequency amplitudes from the Gaussian pulse search and $100$ uniformly spaced durations from $0.1$ to $10$. The Gaussian pulse search and the flat-top pulse search each had $10{,}000$ candidate pulses per source and $10^8$ pairs of pulses from sources A and B. We evaluated the objective function $\mathcal J$ for every Gaussian pulse pair and every flat-top pulse pair using the nonidentical source parameters in Table~\ref{tab:physical_params}. For each pulse shape, we selected the pair with the highest value of $\mathcal J$. Table~\ref{tab:_initializers} lists the selected pulse amplitudes and widths or durations.

\begin{table*}[tbp]
    \centering
    \caption{Parameters of the initial Rabi pulses selected for Table~\ref{tab:main-score-comparison}. Here, $A_\mu$ is the Rabi frequency amplitude, $\sigma_\mu$ is the Gaussian pulse width, and $D_\mu$ is the flat-top pulse duration for source $\mu\in\{A,B\}$. Values are rounded consistently with the main text.}
    \label{tab:_initializers}
    \small
    \begin{tabular}{lcc}
      \toprule
      Pulse shape & Source A & Source B \\
      \midrule
      Gaussian&
      $A_A=1.3667,\ \sigma_A=1.6667$ &
      $A_B=1.1939,\ \sigma_B=1.6667$ \\
      Flat-top&
      $A_A=0.8101,\ D_A=10$ &
      $A_B=0.6949,\ D_B=10$ \\
    \bottomrule
  \end{tabular}
\end{table*}

\subsection{\label{app:optimization_behavior}Optimization Histories and Rabi Pulses}
\begin{figure*}[tbp]
\centering
\includegraphics[width=\textwidth]{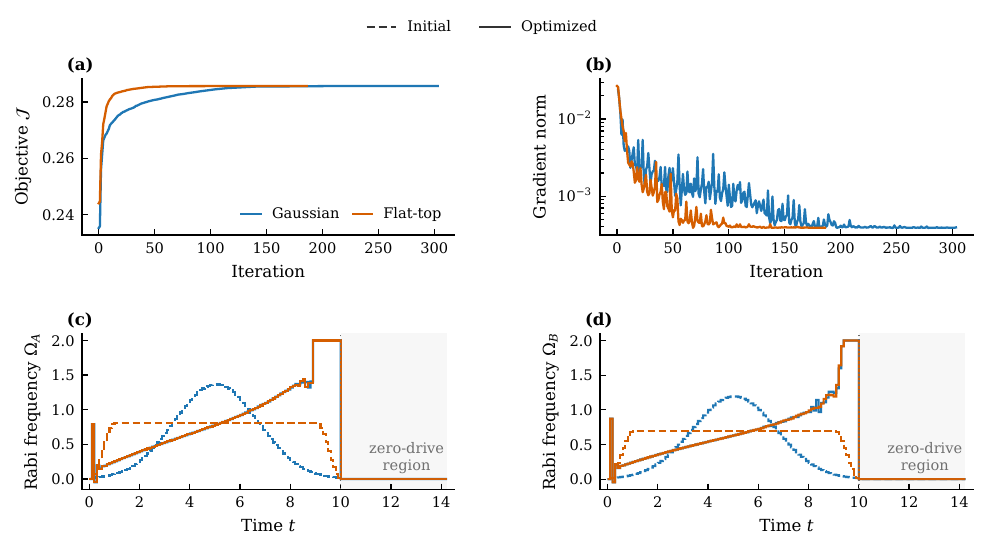}
\caption{QRT-GRAPE optimization histories and Rabi pulses for the Gaussian and flat-top pulse pairs used in Table~\ref{tab:main-score-comparison}.
(a) Objective function value $\mathcal J$.
(b) Euclidean norm of the gradient of $\mathcal J$ with respect to the $198$ adjustable Rabi frequencies.
The horizontal axis in (a,b) counts accepted L-BFGS-B iterations, with $0$ denoting the initial pulses.
(c,d) Initial and optimized Rabi pulses for sources A and B, respectively.
Blue and orange indicate runs starting from Gaussian and flat-top Rabi pulses, respectively; dashed and solid lines in (c,d) denote initial and optimized pulses.
Both atomic detunings are fixed at zero.
The shaded region at $t\geq10$ is the additional decay interval, during which both Rabi frequencies are fixed at zero.}
\label{fig:pulse_shape}
\end{figure*}

Figure~\ref{fig:pulse_shape} shows the QRT-GRAPE optimization histories starting from the Gaussian and flat-top Rabi pulse pairs selected for Table~\ref{tab:main-score-comparison}.

Figure~\ref{fig:pulse_shape}(a) shows that the two selected initial Rabi pulse pairs lead to nearly the same final objective function value.
Starting from the Gaussian Rabi pulse pair, $\mathcal J$ increases from $0.2354$ to $0.2857$ over $303$ accepted iterations.
Starting from the flat-top Rabi pulse pair, it increases from $0.2441$ to $0.2857$ over $186$ accepted iterations.
The optimized Rabi pulses also closely resemble one another for each source, as shown in Fig.~\ref{fig:pulse_shape}(c,d).
In both cases, the optimized Rabi frequencies increase through most of the driving period and reach the upper bound before $t=10$.

As shown in Fig.~\ref{fig:pulse_shape}(b), the Euclidean norm of the gradient of $\mathcal J$ decreases overall, with fluctuations, but remains nonzero at termination.
Several optimized Rabi frequencies reach the upper bound $\Omega_\mu=2$.
Under these amplitude constraints, the Euclidean gradient norm need not vanish at an optimum because directions that would increase $\mathcal J$ can be excluded by the bounds.

The gradual increase in the optimized Rabi frequencies produces pulses that are strongly asymmetric in time, as shown in Fig.~\ref{fig:pulse_shape}(c,d).
Ref.~\cite{Tanji2024Ratefidelity} indicates that asymmetric Gaussian pump pulses with a longer rise than fall could improve the conditional Bell-state fidelity at a given photon-emission probability in their model.
The gradual rise found here may similarly promote population transfer from $\ket{u0}$ to $\ket{g1}$ while limiting the population of $\ket{e0}$ and the associated spontaneous emission.
The switch to zero at $t=10$ is imposed by our control conditions.

\subsection{\label{app:random_rabi}Optimization from Random Initial Rabi Pulses}

We also performed 100 QRT-GRAPE optimizations starting from independently generated random Rabi pulses, with both atomic detunings fixed at zero.
The source pair is specified in Table~\ref{tab:physical_params}.
For each source and each of the $99$ adjustable time intervals, the initial Rabi frequency is picked independently from a uniform distribution on $[-2,2]$.
These initial Rabi frequencies are used directly, without smoothing.
QRT-GRAPE optimizes the $99$ Rabi frequencies of each source within $[-2,2]$, giving $198$ adjustable values in total.

Each trial uses L-BFGS-B with the analytic QRT-GRAPE gradient, at most $500$ iterations, and $\texttt{ftol}=\texttt{gtol}=10^{-8}$.
The trial stops earlier if either convergence criterion is met.
We use the value of $\mathcal J$ at termination in the histogram, including trials that reach the iteration limit.
We count trials with an increase in $\mathcal J$ separately from those for which L-BFGS-B reports convergence.

\begin{figure}[tbp]
\centering\includegraphics[width=\columnwidth]{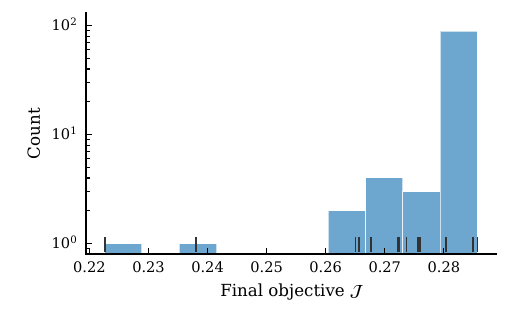}
\caption{Histogram of the objective function value $\mathcal J$ at termination for $100$ QRT-GRAPE optimizations starting from independent random Rabi pulse pairs, with both atomic detunings fixed at zero.
The final values are grouped into $10$ equal-width bins.
The vertical axis is logarithmic, and the short marks along the horizontal axis indicate individual final values.
Each trial runs for at most $500$ L-BFGS-B iterations and stops earlier if a convergence criterion is met.
Trials that reach the iteration limit are included.}
\label{fig:app_random_rabi}
\end{figure}

All $100$ trials ended with a larger objective function value $\mathcal J$ than their respective initial values.
L-BFGS-B reported convergence for $79$ trials, and the remaining $21$ reached the $500$-iteration limit.
Figure~\ref{fig:app_random_rabi} shows the distribution of the final values from all $100$ trials.
The final values ranged from $0.2226$ to $0.2857$, with a median of $0.2857$.
Of the $100$ trials, $87$ ended with $\mathcal J\geq0.285$.

The concentration of final values near $\mathcal J=0.2857$ in Fig.~\ref{fig:app_random_rabi} shows that high objective function values were reached from many different initial Rabi pulse pairs.
The lower-valued tail shows that a single run can terminate with a smaller objective function value under the same stopping conditions.
Running the optimization from several initial Rabi pulse pairs and retaining the largest final value of $\mathcal J$ therefore provides a practical way to reduce dependence on initialization in this example.

\section{\label{app:smoothness_regularization} Smoothing optimized pulses}
The optimized Rabi pulses in Fig.~\ref{fig:pulse_shape} contain steep variations between neighboring time intervals.
To explore smoother pulses for experimental implementation while seeking high objective function values, we included a penalty on these variations during optimization.

All three optimizations started from the Gaussian Rabi pulse pair selected for Table~\ref{tab:main-score-comparison}, with both atomic detunings fixed at zero. For each source, the Rabi frequency in the first time interval was fixed at zero, and QRT-GRAPE adjusted the remaining 99 values within $[-2,2]$, giving 198 adjustable values in total.

Following the general objective function in Eq.~\eqref{eq:J_general}, we maximized the penalized objective function
% The regularized optimization maximizes
\begin{equation}
  \mathcal{J}_{\lambda}[u]
  =
  \mathcal J[u]
  -
  \lambda R[u],
  \label{eq:smooth_penalized_objective}
\end{equation}
where $\mathcal J[u]$ is the objective function defined in the main text and

% We define the roughness of the optimized controls as a measure of changes between neighboring control amplitudes:

\begin{equation}
  R[u]
  =
  \sum_{\mu\in\{A,B\}}
  \sum_{k=1}^{N-1}
  \frac{
  \left(
  u^{(\mu)}_{1,k+1}-u^{(\mu)}_{1,k}
  \right)^2
  }{\Delta t}
  \label{eq:smooth_roughness}
\end{equation}
measures variation between neighboring Rabi frequencies.
Here, $u^{(\mu)}_{1,k}$ is the Rabi frequency for source $\mu$ in time interval $k$, $N=100$, and $\Delta t=0.1$.
The sum includes the change from the first fixed zero value to the next value.
It excludes the change from the final driven interval to the additional decay interval, during which both Rabi frequencies are fixed at zero.

The corresponding gradient used by the optimizer is
\begin{equation}
  \frac{\partial \mathcal{J}_{\lambda}}{\partial u^{(\mu)}_{1,k}}
  =
  \frac{\partial \mathcal{J}}{\partial u^{(\mu)}_{1,k}}
  -
  \lambda
  \frac{\partial R}{\partial u^{(\mu)}_{1,k}}.
  \label{eq:smooth_penalized_gradient}
\end{equation}
The derivative of $\mathcal J$ is evaluated using Eq.~\eqref{eq:QRT_GRAPE_master_gradient}, and the derivative of $R[u]$ is obtained by differentiating Eq.~\eqref{eq:smooth_roughness}.
We compared $\lambda=0$, $10^{-4}$, and $10^{-3}$.

\begin{figure*}[tbp]
  \centering
  \includegraphics[width=\textwidth]{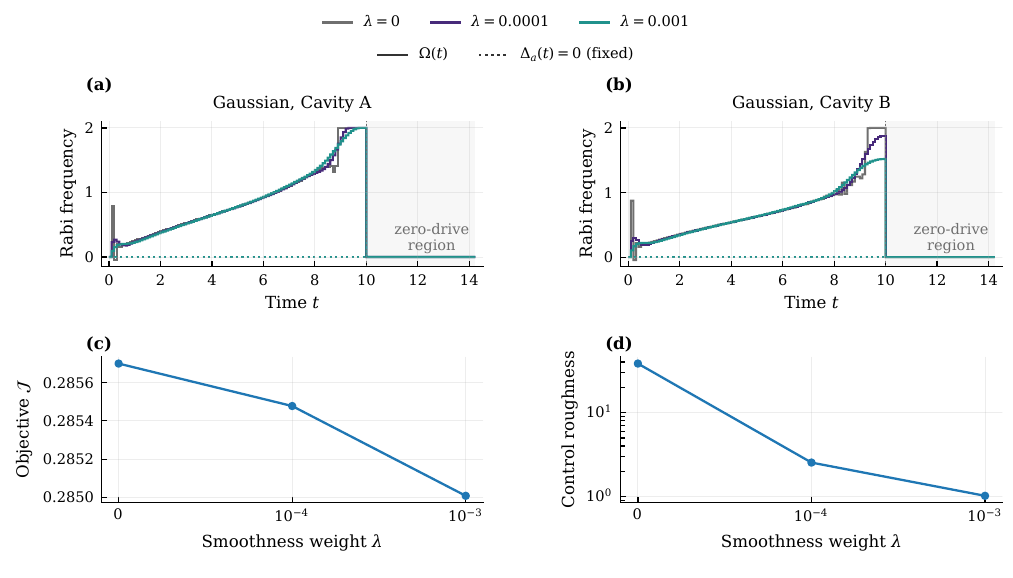}
  \caption{
    Effect of penalizing variations between neighboring Rabi frequencies during optimization.
All three optimizations started from the Gaussian Rabi pulse pair selected for Table~\ref{tab:main-score-comparison}, with both atomic detunings fixed at zero.
(a), (b) Optimized Rabi frequencies for sources A and B, respectively, for $\lambda=0$, $10^{-4}$, and $10^{-3}$.
(c) Objective function value $\mathcal J$, evaluated with the penalty.
(d) Measure $R[u]$ of variation between neighboring Rabi frequencies, shown on a logarithmic scale.
Shaded regions in (a) and (b) indicate the additional decay interval after the Rabi pulses are turned off.
The change in each Rabi frequency when the pulses are turned off is excluded from $R[u]$.
  }
  \label{fig:app_smoothness_comparison}
\end{figure*}
Figure~\ref{fig:app_smoothness_comparison} shows how the smoothness penalty changes the optimized Rabi frequencies and the resulting objective function value $\mathcal J$.

Without the penalty, the optimized pulses gave $\mathcal J=0.2857$ and $R[u]=38.30$.
For $\lambda=10^{-4}$, $R[u]$ decreased to $2.543$, while $\mathcal J$ remained at $0.2855$.
Increasing the coefficient to $\lambda=10^{-3}$ further reduced $R[u]$ to $1.023$, with $\mathcal J=0.2850$.
Relative to $\lambda=0$, the latter pulses reduced $R[u]$ by $97.3\%$, with a $0.242\%$ decrease in $\mathcal J$.
All three calculations satisfied the L-BFGS-B convergence criterion.

Including the penalty thus allowed QRT-GRAPE to find smoother Rabi pulses while retaining nearly the same objective function value.
The coefficient $\lambda$ sets the balance between suppressing variations in the Rabi frequencies and maximizing $\mathcal J$, providing a way to account for pulse smoothness when designing controls for experimental implementation.

\section{\label{app:detuning_initialization}Sensitivity to Initial Detunings}
We examined whether optimizing the atomic detunings together with the Rabi frequencies improved the final objective function value $\mathcal J$. All runs started from the Gaussian Rabi pulse pair selected for Table~\ref{tab:main-score-comparison}, defined in Eq.~\eqref{eq:gaussian_initializer} and listed in Table~\ref{tab:_initializers}. We compared the final objective function values with $\mathcal J_{\mathrm R}$, the existing result obtained by optimizing the Rabi frequencies alone from the same Gaussian pulse pair, with both atomic detunings fixed at zero.

We performed $30$ optimizations in which both Rabi frequencies and both atomic detunings were adjusted. In each run, the initial atomic detuning values for sources A and B in the $99$ adjustable time intervals were drawn independently from a uniform distribution over $[-2,2]$, without smoothing, rescaling, or clipping. All runs used the same initial Gaussian Rabi pulse pair. The value in the first time interval of each function was fixed at zero, and all four functions were set to zero for $t\geq10$. Each run optimized $99$ values of each Rabi frequency and atomic detuning, giving $396$ adjustable values with bounds $[-2,2]$.

\begin{figure}[tbp]
\centering\includegraphics[width=\columnwidth]{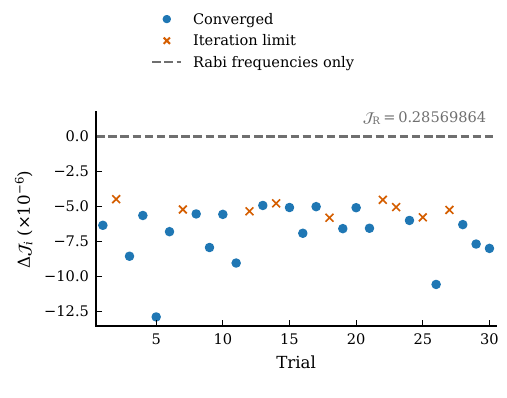}
\caption{Comparison of objective function values from 30 optimizations of both Rabi frequencies and atomic detunings.
All runs started from the same Gaussian Rabi pulse pair, with initial atomic detuning values drawn independently from a uniform distribution over $[-2,2]$ in each adjustable time interval.
The plotted difference is $\Delta\mathcal J_i=\mathcal J_{\mathrm{R+D}}^{(i)}-\mathcal J_{\mathrm R}$, where $\mathcal J_{\mathrm{R+D}}^{(i)}$ is the objective function value at termination and $\mathcal J_{\mathrm R}$ is the existing result obtained by optimizing the Rabi frequencies alone from the same Gaussian pulse pair, with both atomic detunings fixed at zero.
The dashed line marks zero difference and is annotated with $\mathcal J_{\mathrm R}=0.28569864$.
Circles denote the 21 runs for which L-BFGS-B reported convergence; crosses denote the nine runs that reached the 500-iteration limit.}
\label{fig:app_detuning}
\end{figure}

For run $i$, we define the change from this existing result as
\begin{equation}
  \Delta\mathcal J_i
  =
  \mathcal J_{\mathrm{R+D}}^{(i)}
  -
  \mathcal J_{\mathrm R}.
\end{equation}
Figure~\ref{fig:app_detuning} shows the differences for all 30 runs.

None of the 30 optimizations of both Rabi frequencies and atomic detunings exceeded the existing result obtained by optimizing the Rabi frequencies alone.
The differences $\Delta\mathcal J_i$ ranged from $-1.2912\times10^{-5}$ to $-4.4877\times10^{-6}$, with a median of $-5.8007\times10^{-6}$.
These values include all 30 runs, including the nine that reached the iteration limit.

The existing solution obtained by optimizing the Rabi frequencies alone, with both atomic detunings fixed at zero, remains feasible when both the Rabi frequencies and atomic detunings are optimized.
Adding atomic detunings as optimization variables therefore cannot lower the maximum attainable objective function value.
In all 30 runs, we optimized both the Rabi frequencies and atomic detunings, starting from the initial Gaussian Rabi pulse pair and random initial atomic detuning values.
The previously optimized Rabi pulses were not used as the initial values.
These different initial conditions may have led the optimization along paths that did not reach the existing objective function value within the iteration limit and stopping criteria used here.
Since optimizing both the Rabi frequencies and atomic detunings provided no improvement in this comparison, we fixed both atomic detunings at zero for simplicity in the numerical examples in the main text.

\section{\label{app:gradient_verification}Numerical Verification of the Analytic Gradient}

We evaluated the gradient for one fixed set of Rabi frequencies and atomic detunings.
The physical parameters were those listed in Table~\ref{tab:physical_params}.
For the two sources, we collect the control values defined in the main text as
\begin{equation}
\begin{gathered}\bm u=\{u^{(\mu)}_{m,k}\},\qquad \mu\in\{A,B\},\\ m\in\{1,2\},\quad k=1,\ldots,100.\end{gathered}
\end{equation}
Here, $u^{(\mu)}_{1,k}$ and $u^{(\mu)}_{2,k}$ are the Rabi frequency and atomic detuning, respectively, for source $\mu$ in time interval $k$.
Each function is represented by $100$ values during the driving period $0\leq t<10$, with one value held constant over each interval of width $\Delta t=0.1$.
In the remaining $99$ intervals, the Rabi frequencies were fixed at $\Omega_A=1.0$ and $\Omega_B=0.8$, and both atomic detunings were fixed at $0.01$.
Both Rabi frequencies and both atomic detunings were fixed at zero during the additional decay interval, which was divided into $10$ intervals.

For each selected Rabi frequency or atomic detuning value $u^{(\mu)}_{m,k}$, we compared the analytic derivative $g^{\mathrm{ana},(\mu)}_{m,k}=\partial\mathcal J/\partial u^{(\mu)}_{m,k}$ with the centered finite-difference estimate
\begin{equation}
g^{\mathrm{FD},(\mu)}_{m,k}(\epsilon)
=\frac{\mathcal J(\bm u+\epsilon\bm e^{(\mu)}_{m,k})-\mathcal J(\bm u-\epsilon\bm e^{(\mu)}_{m,k})}{2\epsilon}.
\end{equation}
Here, $\bm u$ denotes the vector of control values, and $\bm e^{(\mu)}_{m,k}$ denotes the standard basis vector whose component corresponding to $u^{(\mu)}_{m,k}$ is one and whose other components are all zero. Thus, the finite difference changes only that control value.
We selected the time intervals
\begin{equation}
k\in\{2,26,51,76,100\}
\end{equation}
for each Rabi frequency and atomic detuning of sources A and B, giving $20$ derivatives to compare in total.
The values in the first interval ($k=1$) and the additional decay interval were fixed at zero and were not varied in the finite-difference calculation.

\begin{figure*}[tbp] \centering \begin{minipage}[t]{0.49\textwidth} \vspace{0pt} \includegraphics[width=\linewidth]{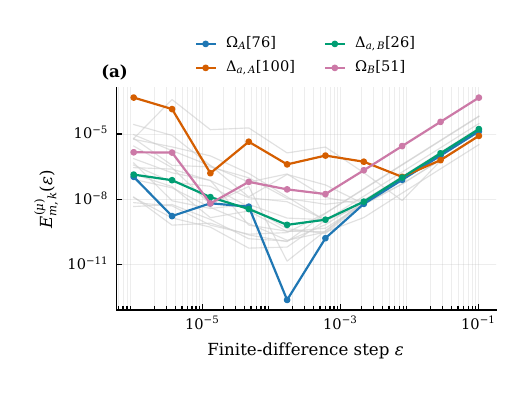} \end{minipage}\hfill \begin{minipage}[t]{0.49\textwidth} \vspace{0pt} \IfFileExists{figs/appendix_gradient_timing.pdf}{%
  \includegraphics[width=\linewidth]{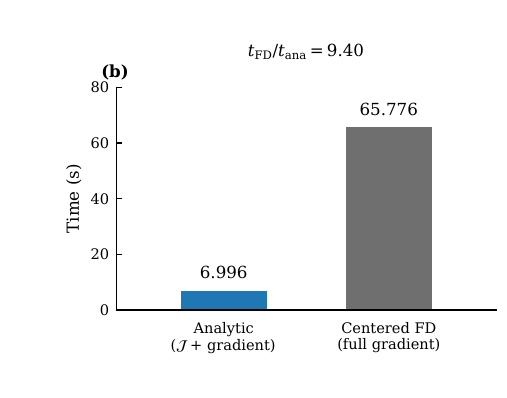}%
  }{%
    \fbox{\begin{minipage}[c][2.65in][c]{\dimexpr\linewidth-2\fboxsep-2\fboxrule\relax}   \centering   (b)\\[1em]   \textbf{[VALUE: TIMING PLOT]}\\[1em]   Analytic / Centered FD\\   Time (s)   \end{minipage}}%
    } \end{minipage} \caption{ Numerical verification of the analytic QRT-GRAPE gradient.  (a) Symmetric relative error between the analytic and centered finite-difference derivatives with respect to $20$ selected Rabi frequency and atomic detuning values, as a function of the step size $\epsilon$.  Gray curves show all $20$ selected derivatives, and colored curves highlight four of them.  The values in square brackets in the legend are the time interval indices $k$.  (b) Median wall-clock time for evaluating $\mathcal J$ and all $396$ analytic derivatives, compared with that for computing all $396$ derivatives by centered finite differences.  Both methods were timed on the same computer with the same allocated CPU budget.  } \label{fig:app_fd_check} \end{figure*}

We quantified the difference between the analytic and finite-difference derivatives using the symmetric relative error
\begin{equation}
E^{(\mu)}_{m,k}(\epsilon)
=\frac{\left|g^{\mathrm{FD},(\mu)}_{m,k}(\epsilon)-g^{\mathrm{ana},(\mu)}_{m,k}\right|}{\max\left(\left|g^{\mathrm{FD},(\mu)}_{m,k}(\epsilon)\right|,\left|g^{\mathrm{ana},(\mu)}_{m,k}\right|,10^{-14}\right)}.
\end{equation}
This measures the absolute difference relative to the larger magnitude of the two derivatives and is unchanged when the two derivative values are exchanged.
The lower bound $10^{-14}$ on the denominator prevents division by zero or by a very small value.
We used $10$ values of $\epsilon$, equally spaced on a logarithmic scale within $10^{-6}\leq\epsilon\leq10^{-1}$.
Figure~\ref{fig:app_fd_check}(a) shows the resulting errors for the $20$ selected derivatives.

For each derivative, we first took the smallest error among the $10$ tested step sizes:
\begin{equation}
E^{(\mu)}_{m,k,\min}=\min_{\epsilon\in\mathcal E}E^{(\mu)}_{m,k}(\epsilon),
\end{equation}
where $\mathcal E$ is the set of tested step sizes.
The step size giving the smallest error differed among the derivatives.
We then took the largest of these $20$ minimum errors:
\begin{equation}
\max_{\mu,m,k}E^{(\mu)}_{m,k,\min}=1.0713\times10^{-7},
\end{equation}
where the maximum is taken over the $20$ selected derivatives.
The agreement shown in Fig.~\ref{fig:app_fd_check}(a) confirms that our implementation correctly evaluates the analytic derivatives of $\mathcal J$ with respect to the $20$ tested Rabi frequency and atomic detuning values.

We compared the time required to evaluate $\mathcal J$ and all $396$ analytic derivatives with that required to compute the $396$ derivatives by centered finite differences.
Both calculations used the same fixed Rabi frequencies and atomic detunings, physical parameters, time grid, and trapezoidal integration.
Centered finite differences required $792$ evaluations of $\mathcal J$ at one fixed step size $\epsilon=10^{-4}$ .
Both methods ran sequentially on Intel Xeon Silver 4410Y system running Linux under WSL2 in Docker, with the same allocated CPU budget of one logical CPU.
For each method, we used one process, with one thread per process for matrix calculations.
The software versions were Python 3.10.12, NumPy 1.26.4~\cite{harris2020array}, SciPy 1.15.1~\cite{2020SciPy-NMeth}, OpenBLAS 0.3.23.dev for NumPy, and OpenBLAS 0.3.28 for SciPy~\cite{OpenBLAS}.
We first evaluated $\mathcal J$ and its analytic gradient once and computed the full gradient by centered finite differences once.
We then timed five further evaluations with each method and report the median time, excluding the initial evaluation.
The timed operations were evaluation of $\mathcal J$ and its analytic gradient for the analytic calculation, and all $792$ objective function evaluations and assembly of the derivatives for centered finite differences.

Figure~\ref{fig:app_fd_check}(b) compares the median times for the two calculations.
The median time was $6.996$\,s for evaluating $\mathcal J$ and its analytic gradient and $65.776$\,s for computing the gradient by centered finite differences.
For this input, evaluating $\mathcal J$ and its analytic gradient was 9.40 times as fast as computing the gradient by centered finite differences.
This reduces the time spent evaluating gradients during pulse optimization.

\bibliography{ref}

\end{document}